\documentclass[14pt, preprint]{aastex62}

\def\lesssim{\mathrel{\hbox{\rlap{\hbox{\lower4pt\hbox{$\sim$}}}\hbox{$<$}}}}
\def\gtrsim{\mathrel{\hbox{\rlap{\hbox{\lower4pt\hbox{$\sim$}}}\hbox{$>$}}}}

\usepackage{graphicx}
\usepackage{float}

\def\so{\Sigma_{\rm 0}}
\def\B{\begin{equation}}
\def\E{\end{equation}}

\def\B{\begin{equation}}
\def\E{\end{equation}}
\def\O{\Omega}

\def\ni{\noindent}

\def\D{\Delta}

\def\ep{\epsilon}

\shorttitle{Effects of Dissipation Profiles on Disk Spectra}
\shortauthors{Dezen, Egger \& Mwansa}

\begin{document}

\title{Effects of Strong Photospheric Dissipation on the Spectra and Structure of Accretion Disks with Non-zero Inner Torque}

\author{Theodore Dezen}
\email{tdezen@sandiego.edu}
\affil{Department of Physics and Biophysics, University of San Diego, San Diego, CA 92110}
\author{Noah Egger}
\affil{Department of Physics, San Diego State University, San Diego, CA 92182}
\author{Lwendo Mwansa}
\affil{Department of Physics and Biophysics, University of San Diego, San Diego, CA 92110}


\begin{abstract}

\ni We present numerical calculations of spectra and structure of accretion disks models appropriate for near-Eddington luminosity black hole X-ray binaries (BHB). Our work incorporates non-zero torque at the ISCO as well as several dissipation profiles based on first-principles three-dimensional disk interior simulations. We found that significant dissipation near the photosphere can produce steep power law-like spectra for models with moderate viewing angles spanning a range of black hole spins while including inner torque push the spectral peak to higher energies. Consistent with previous studies, we also conclude that disks with stresses at the inner edge remain viable models for high-frequency quasi-periodic oscillations (HFQPO), especially given that increasing dissipation near the photospheres actually resulted in QPO power spectra with higher quality factors compared to those found in recent work.

\end{abstract}

\section{Introduction}

Black hole X-ray binaries (BHBs) in the Milky Way typically display several outburst states over a wide range of luminosities that correlate with different spectral shapes and variability properties \citep{mr06, dgk07}. While there has been tremendous recent progress in understanding the accretion processes that presumably power these systems, the internal structure of the flow at the highest observed luminosities remain an open problem. Specifically, the spectra of BHBs radiating near the Eddington limit may contain a broad steep power law (SPL, photon index $\Gamma>2.4$) component that begins at the spectral peak ($\la10 \ \rm keV$) and can stretch well into gamma ray regime at hundreds of keVs or even MeVs \citep{gr98, lw05}. While in the SPL state, BHBs also display high-frequency quasi-periodic oscillations (HFQPO, $\nu>50 \ \rm Hz$) in their hard X-ray light curves in the approximately $10$ to $30 \ \rm keV$ band. 

\

The advent of vertically stratified, three-dimensional shearing box simulations led to important advances in understanding how the gravitational potential energy lost during accretion is dissipated via MRI turbulence \citep{bh91, bh98} and transported within the flow. Over a wide range of box-integrated radiation to gas pressures \citep{tur04, hir06, kro07, bla07, hkb09}, the spatial dissipation rates generally peak at about a pressure scale height above (or below) the mid-plane \citep{bla11}. More recent global simulations \citep{sc13, jia14} also found that maximum dissipation occurs at higher altitudes to heat the photospheric region, resulting in a hot corona above the disk that can plausibly lead to SPL-like spectra. 
 
\

While there has been significant progress in integrating more sophisticated radiative transfer schemes into simulations \citep{dav12, jia12}, calculations that self-consistently include full multi-frequency treatments do not yet exist. To connect to observations, many authors have instead incorporated time and horizontally averaged dissipation profiles from the simulations into one-dimensional numerical calculations that couple disk vertical structure and radiative transfer (see for examples, \cite{dav05}, \cite{bla06} and \cite{dav09}). More recently, \cite{tb13} found that stronger dissipation near the disk photosphere may result in a Compton up-scattering corona that leads to SPL-like non-thermal spectral tails. 

\

In another attempt to explain the SPL state, \cite{db14} proposed that in a disk with non-zero magnetic stresses at the inner edge \citep{ak00}, the rapidly inward increasing effective temperature would ensure that the annuli spectral peaks would become further separated for the same radial distance change closer to the black hole and add up to SPL-like full disk spectra. Moreover, such a scenario would naturally explain why HFQPO power spectra peaks become narrower when restricted to integrating over higher energy bands (approximately $> 10 \ \rm keV$) since these photons primarily come from a narrow radii range in the inner disk. More detailed radiative transfer calculations by \cite{df18} suggested that incorporating inner torque alone is a promising model for HFQPO but may be insufficient to drive the SPL state.

\

In this work, we explore the possibility that strong dissipation near the photosphere combined with a hotter (compared to standard thin disks) inner disk with non-zero torque at the ISCO could lead to both SPL-like spectra and high quality factor HFQPOs. Compared to previous efforts \citep{dav05, tb13, df18}, we significantly expand the parameter space by constructing disk models of near-Eddington accretion flows at multiple black hole spins as well as utilizing several simulation-motivated dissipation profiles. At each disk annulus, our treatment self-consistently couple radiative transfer and vertical structure equations. The resulting spectra fully accounts for effects of metal opacities and Comptonisation. In addition to shedding light on the dissipation process, the shearing box simulations referenced earlier also collectively suggest that the $\alpha$-model \citep{ss73} relationship between pressure and total stress approximately hold \citep{bla11}. We therefore use $\alpha$-model equations modified to include non-zero inner torque to compute the radial disk structure solutions necessary for our spectral calculations. 

\

We begin this paper by outlining our methodology and assumptions in Section $2$, paying particular attention to our treatment of local dissipation rates. We then present numerical results in Section $3$. Finally, we summarize and discuss potential future work in Section $4$.

\section{Methods}

We divide an accretion disk around a $7$ solar mass black hole into approximately $50$ annuli with the inner most ring at the ISCO. For each annulus, we self-consistently compute the vertical structure and local emergent spectra using the one-dimensional stellar atmosphere code TLUSTY \citep{hl95} modified for accretion disk applications \citep{h98, h00}. TLUSTY assumes a stationary plane-parallel atmosphere so that physical quantities such as density and temperature only depend on height above the mid-plane. The code solves the equations of hydrostatic equilibrium, energy and mass conservation, statistical equilibrium of level populations together with that of full multi-frequency radiative transfer. In particular, we do not assume the spectra are color corrected blackbodies, though for spectral fitting purposes it is often useful to extract a spectral hardening factor from TLUSTY calculations \citep{dav18}. We treat electron scattering for both the Thomson and Compton cases via the Kompaneets approximation and explicitly compute ion populations for hydrogen, helium and metals without assuming local thermodynamic equilibrium. Finally, TLUSTY fully incorporates the effects of general relativity in vertical structure equations.

\

\begin{figure}
\includegraphics[width=14cm]{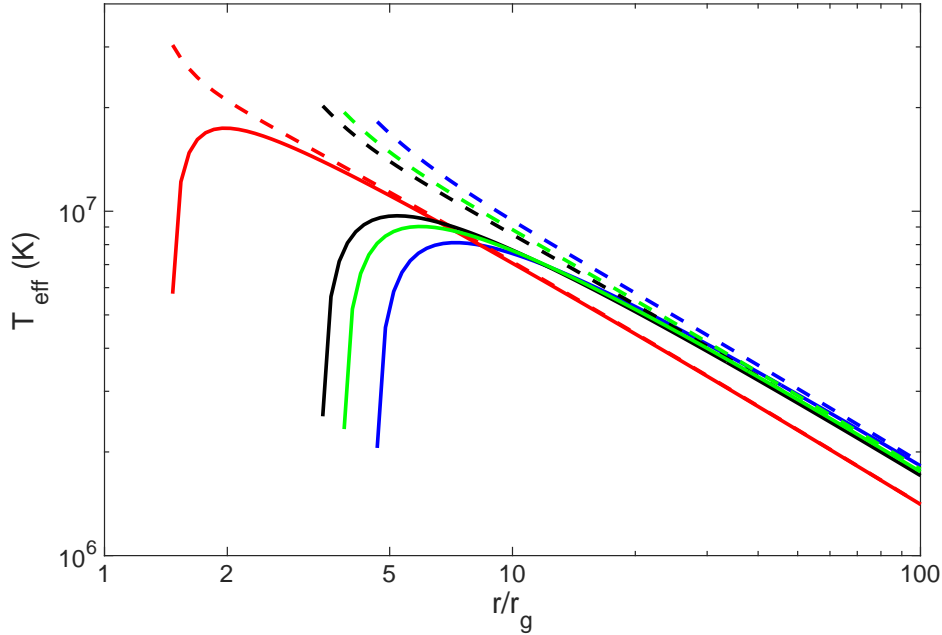}
\caption{Representative radial effective temperature profiles as a function of distance from the black hole normalized to the gravitational radii for accretion disks with zero (solid) and non-zero inner torque (dotted), respectively. The colors correspond to models with different black hole spins: blue, $a/M=0.4$; green, $a/M=0.6$; black, $a/M=0.7$; red, $a/M=0.99$. Note that the effective temperature continues to rise sharply towards the black hole all the way down to the inner edge for the black curve.}
\label{fig:profiles}
\end{figure}

\begin{figure}
\includegraphics[width=14cm]{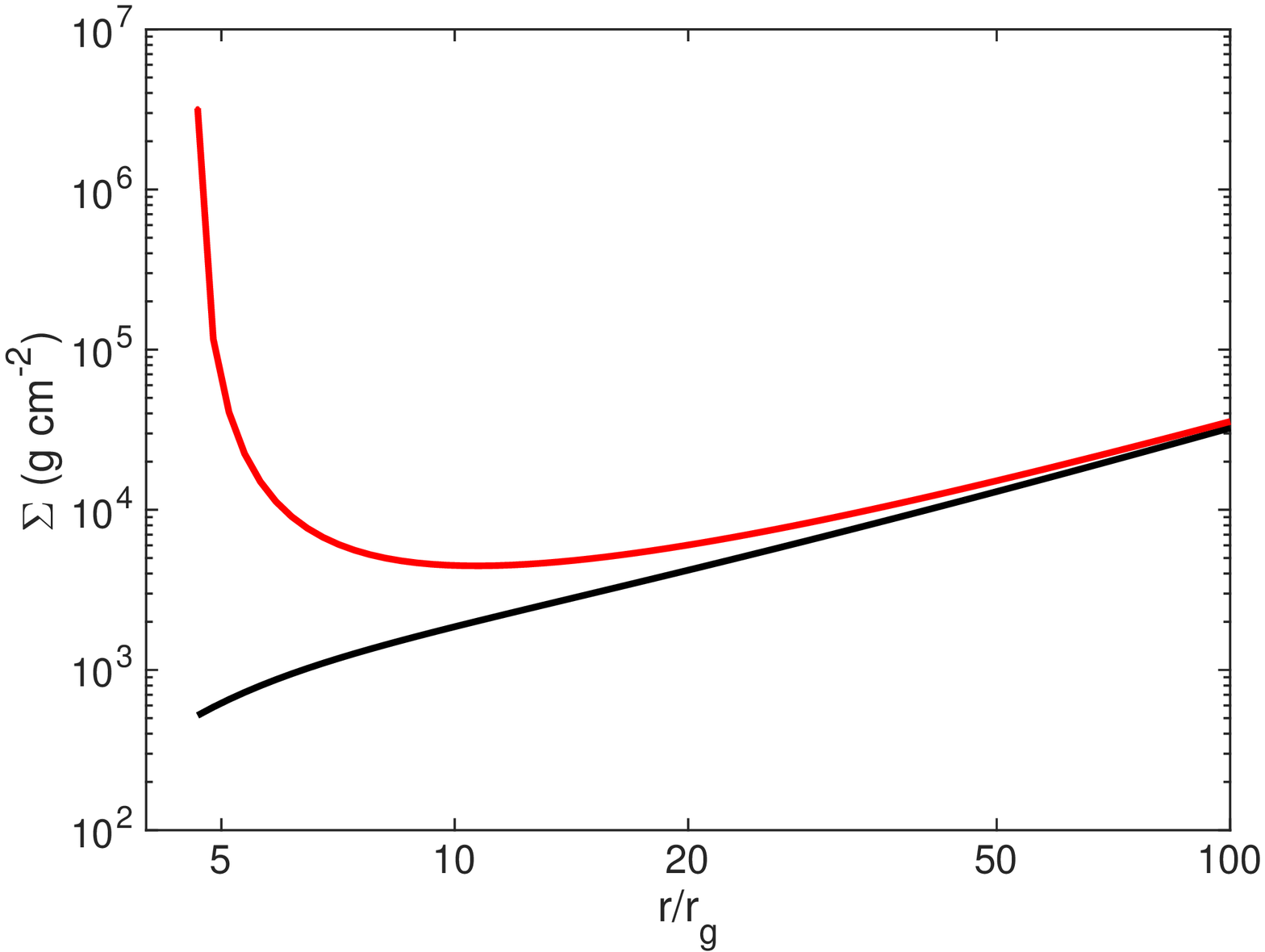}
\caption{Total disk surface density $\Sigma_0$ as a function of distance from black hole for $a/M=0.4$. Note that surface density continues to decrease towards the ISCO when $\Delta\ep=0.1$ (black) but rises sharply for the disk with $\Delta\ep=0$ (red).}
\label{fig:sigma}
\end{figure}

For each annulus, TLUSTY needs the effective temperature $T_{\rm eff}$, total surface density $\Sigma_0$ and square of angular frequency $\O^2$ as input, all of which depends on distance from the black hole which we scale to the gravitational radius $r_g$. We computed these parameters using methods outlined in \cite{ak00} and \cite{db14}, which parametrizes the effects magnetic stresses at the inner disk edge by defining an efficiency enhancement factor $\Delta\ep$ so that the normalized accretion rate is
\B
\dot{m}=\frac{\ep\dot{M}c^2}{L_{\rm Edd}}=\frac{(\ep_0+\Delta\ep)\dot{M}c^2}{L_{\rm Edd}},
\E 
where $M$ is the black hole mass and $\ep_0$ is the radiative efficiency with no torque inner boundary condition. In this work, we choose $\Delta\ep=0.1$ and the resulting luminosity is $L\approx L_{\rm edd}$. As Figure (\ref{fig:profiles}) illustrates, compared to a standard thin disk \citep{ss73, nt73}, our $T_{\rm eff}$ rises sharply towards the black hole instead of drop to zero. In Figure (\ref{fig:sigma}), the effects of inner torque are also visible in the surface density profile, which decreases rather than increases towards the black hole at small radii (within about $10r_g$).

\

\begin{figure}
\includegraphics[width=14cm]{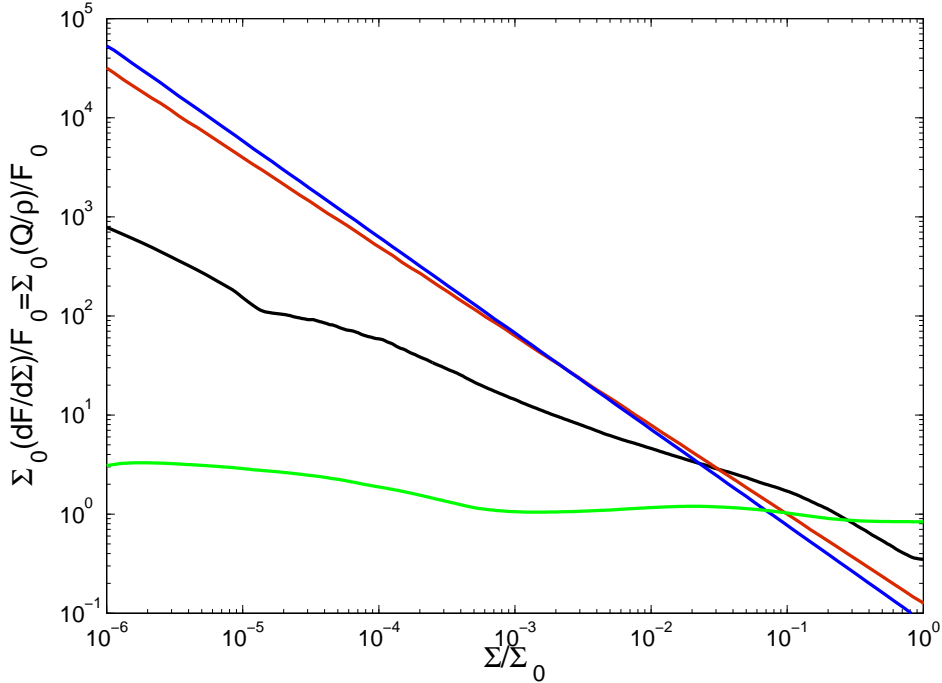}
\caption{Representative dissipation per unit mass normalized to total annulus flux times mid-plane surface density as a function of fractional surface density. The blue and red curves represent the profiles of equation (\ref{dis2}) with $\zeta=0.03$ and $\zeta=0.1$, respectively. For comparison, the black curve is the profile from the \cite{hbk09} simulations and green curve is based on the $\alpha$-prescription, where we took $Q=-\tau_{\rm r\phi}d\Omega/d{{\rm ln} \ r}$ from horizontally and time averaged vertical structure data the same simulation, and used a vertically averaged $\alpha\approx 0.02$. Here we plot $Q/\rho$ instead of $Q\Sigma/\rho$.}
\label{fig:disp}
\end{figure}

In addition to basic annuli parameters, we also specify height-dependent dissipation rates that capture how the gravitational potential energy lost by the accreting material is ultimately converted to thermal energy that heats the plasma. For comparison, we include in this analysis results from our previous paper \citep{df18}, which used a broken power-law fit (black curve Figure \ref{fig:disp}) to the time and horizontally averaged dissipation per unit mass from the shearing box simulations of \cite{hkb09}, namely
\B
\frac{Q\Sigma}{\rho}=-\Sigma\frac{dF}{d\Sigma}=F_0\left\{\begin{array}{ccc}A\left(\frac{\Sigma}{\so}\right)^{0.5}&,& \Sigma/\so<0.11\\B\left(\frac{\Sigma}{\so}\right)^{0.2}&,& \Sigma/\so>0.11\end{array}\right..
\label{dis1}
\E
Here $\rho$ is the mass density, and the surface density $\Sigma$ at height $z$ is defined such that $\Sigma(z\rightarrow\infty)\rightarrow0$ and $z=0$ being the disk mid-plane so that $\Sigma_0=\Sigma(z=0)$. The coefficients $A\approx 0.65$ and $B\approx 0.33$ are normalization constants to ensure that the profile is continuous at the break and the total flux at the disk upper surface is $F_0$. 

\

Our primary dissipation profiles of interest have the form \citep{tb13}
\B
\frac{Q\Sigma}{\rho}=-\Sigma\frac{dF}{d\Sigma}=F_0\zeta\left(\frac{\Sigma}{\so}\right)^{\zeta},
\label{dis2}
\E
which are also power-laws in terms of fractional surface density. In this work, we chose $\zeta=0.1$ or $0.03$. Numerical experiments showed that $0.1$ is the highest value that gave rise to energetically significant non-thermal tails. On the other hand, for $\zeta<0.03$ our code would not capture all of the expected radiative flux because the minimum surface density of the computational domain is limited by convergence considerations. As shown in Figure \ref{fig:disp}, the profile with $\zeta=0.03$ results in higher dissipation in the disk surface layers than the one with $\zeta=0.1$ while both dissipate significantly more power near the photosphere than the one presented in Equation (\ref{dis1}). While our dissipation profiles differ in the fraction of power injected to the disk upper layers, they have peaks away from the disk mid-plane when plotted as a function of height (Figure \ref{fig:qz}), in agreement with findings from recent disk interior simulations. To proceed, we calculate annuli spectra and vertical structure for disks with various dissipation profiles and black hole spins. Our focus is on the disk upper layers where Comptonisation can significantly alter the local spectra. Next, we constructed full disk spectra as seen by distant observers using a relativistic transfer function \citep{agol97}.

\

In computing the vertical structure and resulting spectra, we describe the plasma with a single gas temperature $T_{\rm g}$ since the disk scale height is much greater than electron-ion collision mean free path. Next, we relate the dissipation rate per unit volume $Q$ to the total (that is, frequency-integrated) radiative flux at any height via
\B
Q(z)=\frac{dF}{dz},
\label{qdfdz}
\E
which means vertical radiative diffusion is the only energy transport mechanism we consider. Finally, we also ignore magnetic fields in this study. We refer readers to our previous paper \citep{df18} and references therein for more thorough discussions regarding the justifications and consequences of the latter two assumptions.

\

Before moving on, we point out that radiation pressure dominated systems radiating at near or higher than Eddington limit are likely better described by slim \citep{ab88} instead of thin disks. In these systems both radial and vertical energy transport are important, solving the coupled two-dimensional problem results in radial dependence \citep{sa11} of $\Sigma_0$ and $T_{\rm eff}$ that can differ significantly from that of thin disks. While such models do not yet self-consistently include frequency-dependent radiative transfer, \cite{sdm13} used TLUSTY models \citep{dav05} that corresponded radial quantities from \cite{sa11} to fit observations and found that slim and thin disks only produce slightly different spectra even at near Eddington luminosities, at least in calculations without additional dissipation near the photospheres. We therefore believe that thin disk models are sufficient for qualitatively investigating the effects local dissipation physics on flows with inner magnetic stresses. 

\section{Numerical Results}

\subsection{Corona and Annuli Spectra}

\begin{figure}
\includegraphics[width=9.5cm]{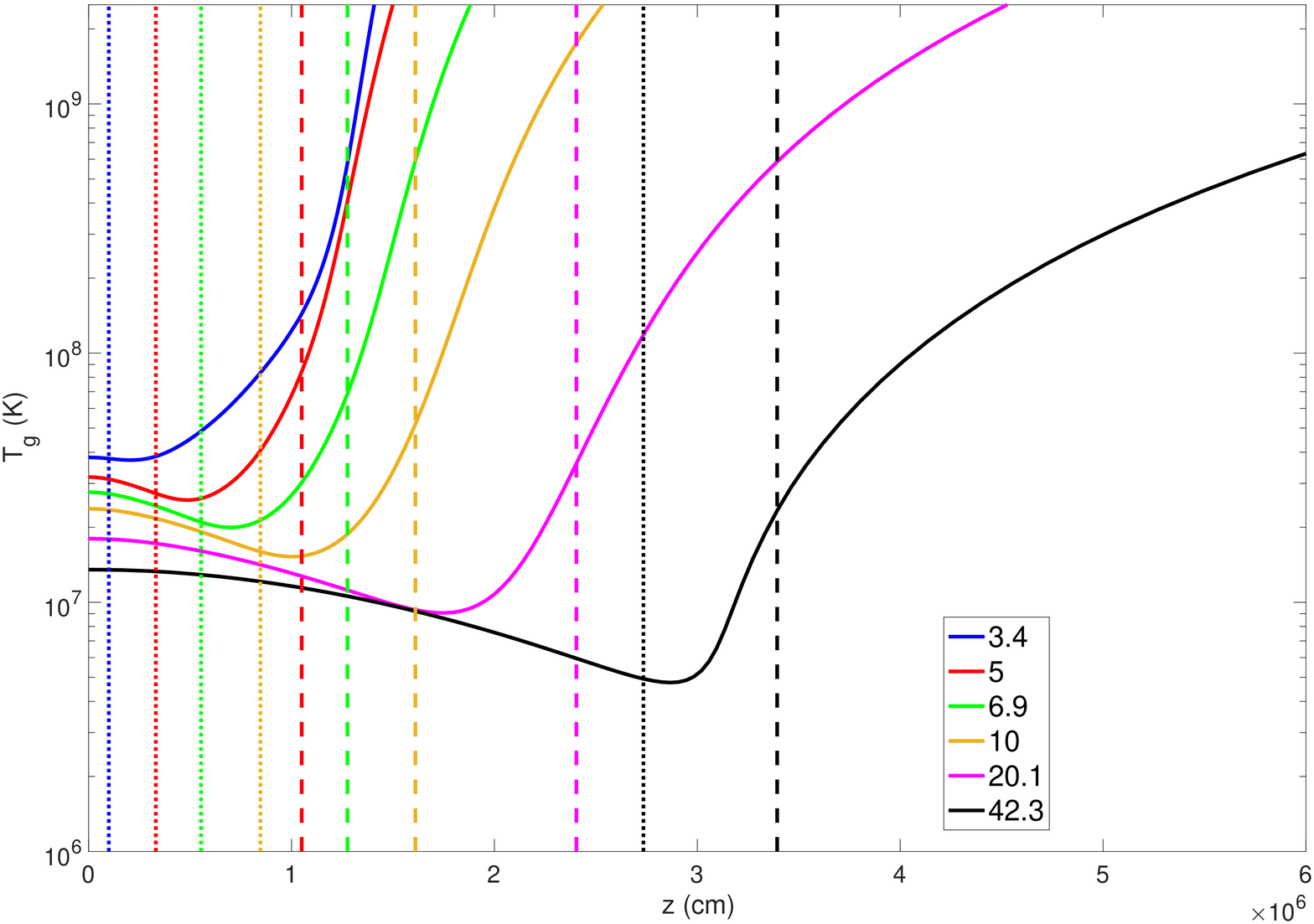}
\includegraphics[width=9.5cm]{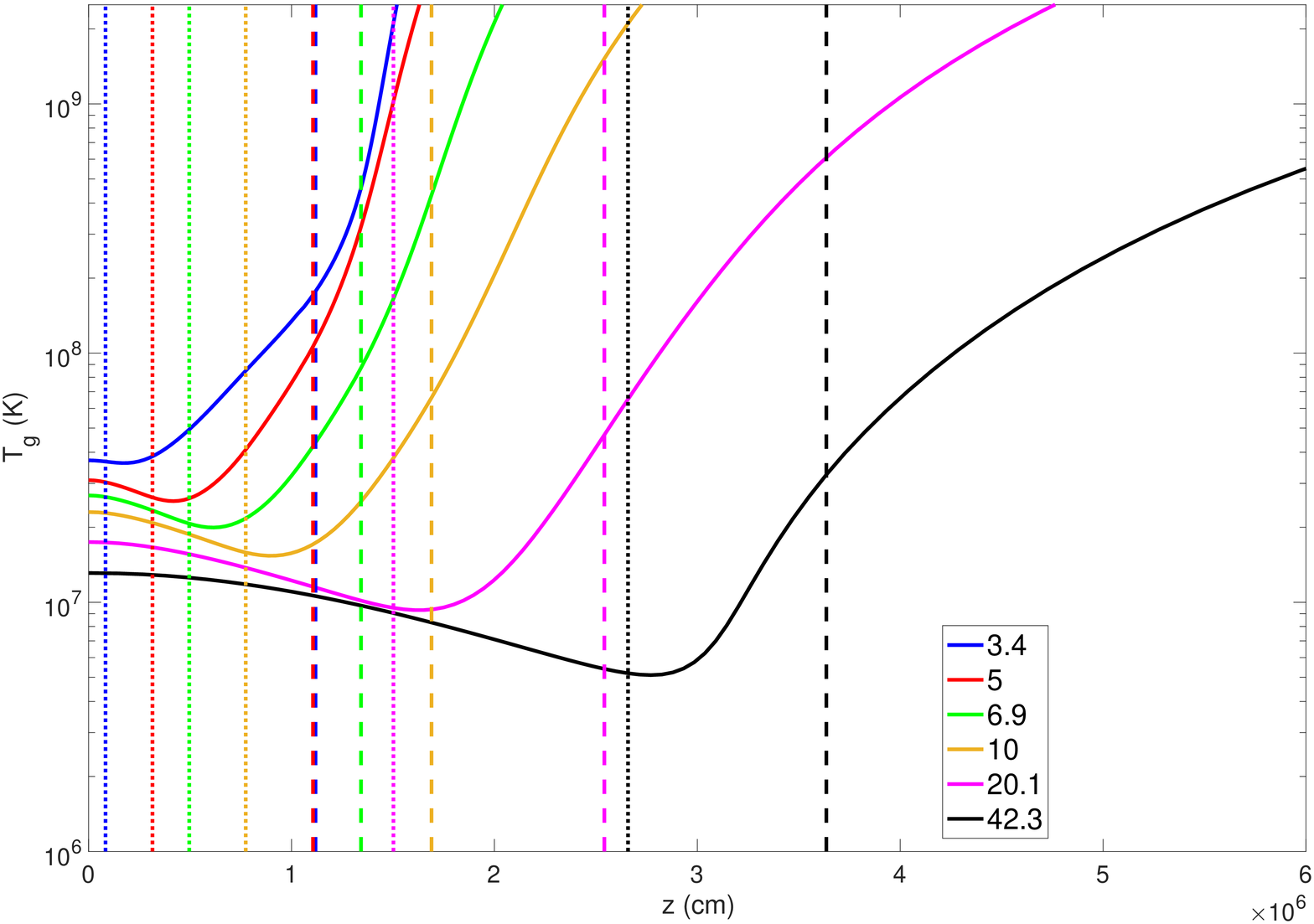}
\includegraphics[width=9.5cm]{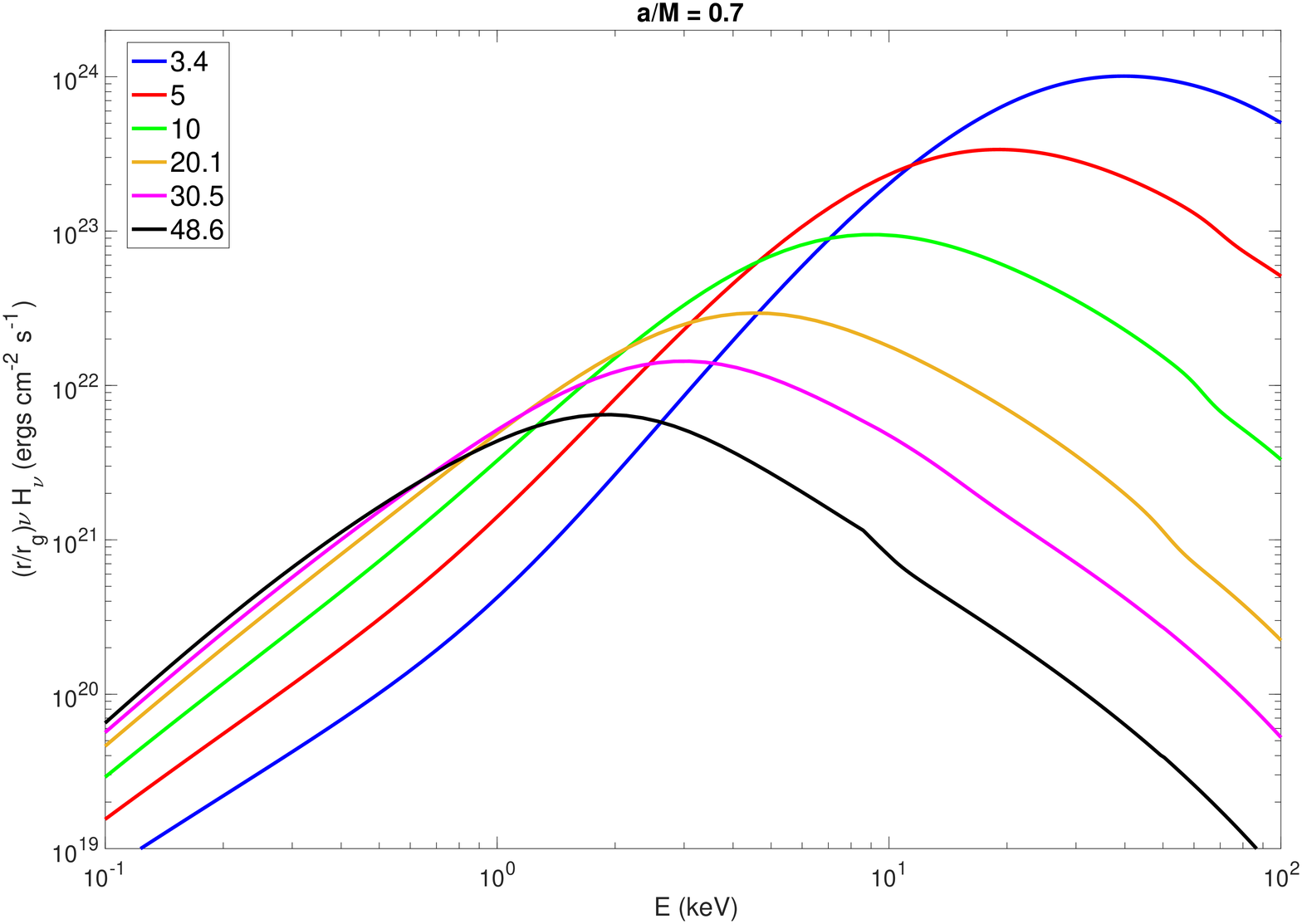}
\includegraphics[width=9.5cm]{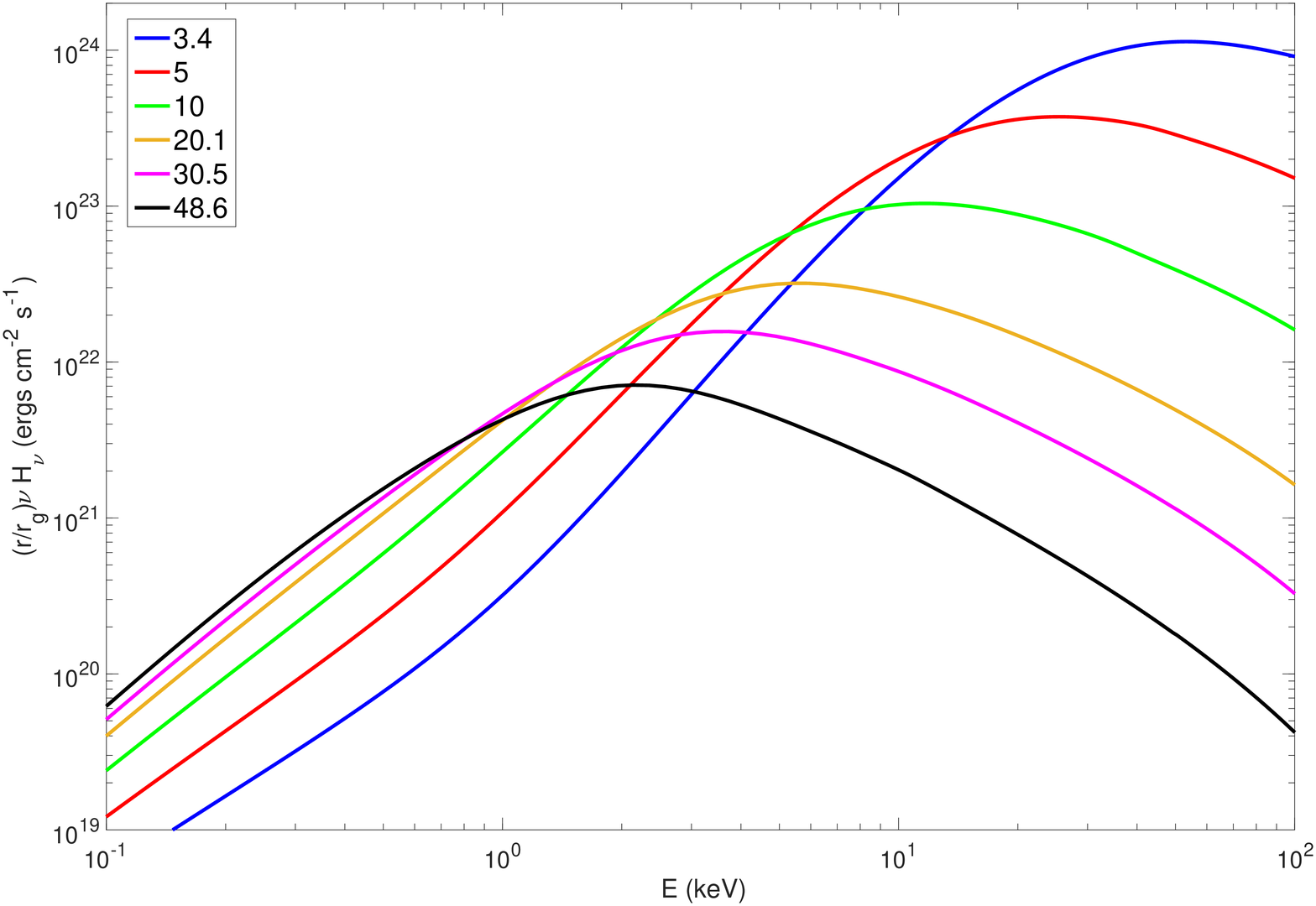}
\caption{Sample gas temperature as functions of height (upper row) above mid-plane for representative annuli of the disk the $a/M=0.7$ black hole and corresponding area-weighted annuli spectra (lower row). The left and right columns represent the $\zeta=0.03$ and $\zeta=0.1$ dissipation profiles, respectively. The dotted and dashed vertical lines in the temperature plots indicate locations of the effective and scattering photospheres, respectively. In all graphs, the legends report distance from black hole, $r/r_g$.}
\label{fig:a7t}
\end{figure}

\begin{figure}
\includegraphics[width=9.5cm]{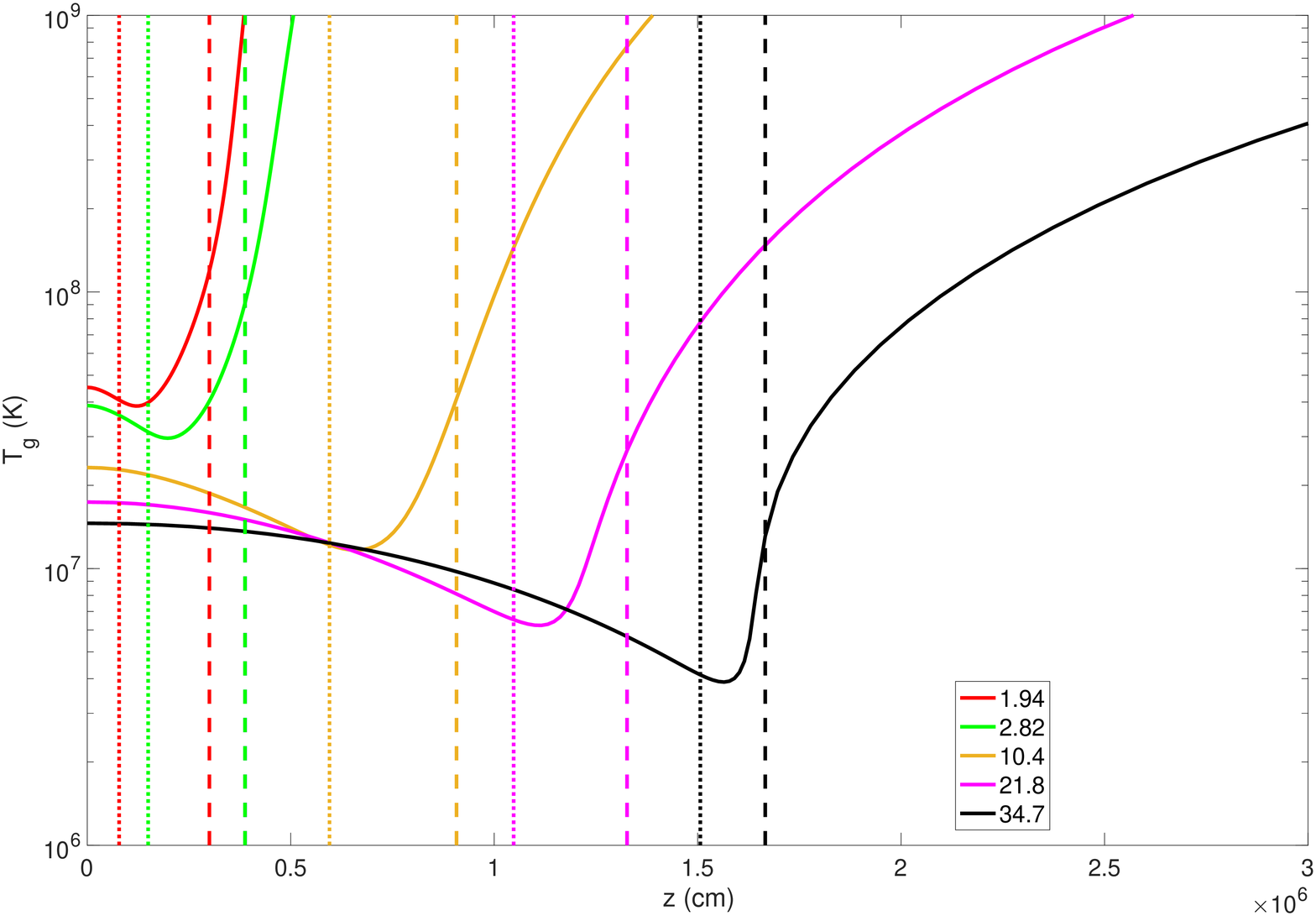}
\includegraphics[width=9.5cm]{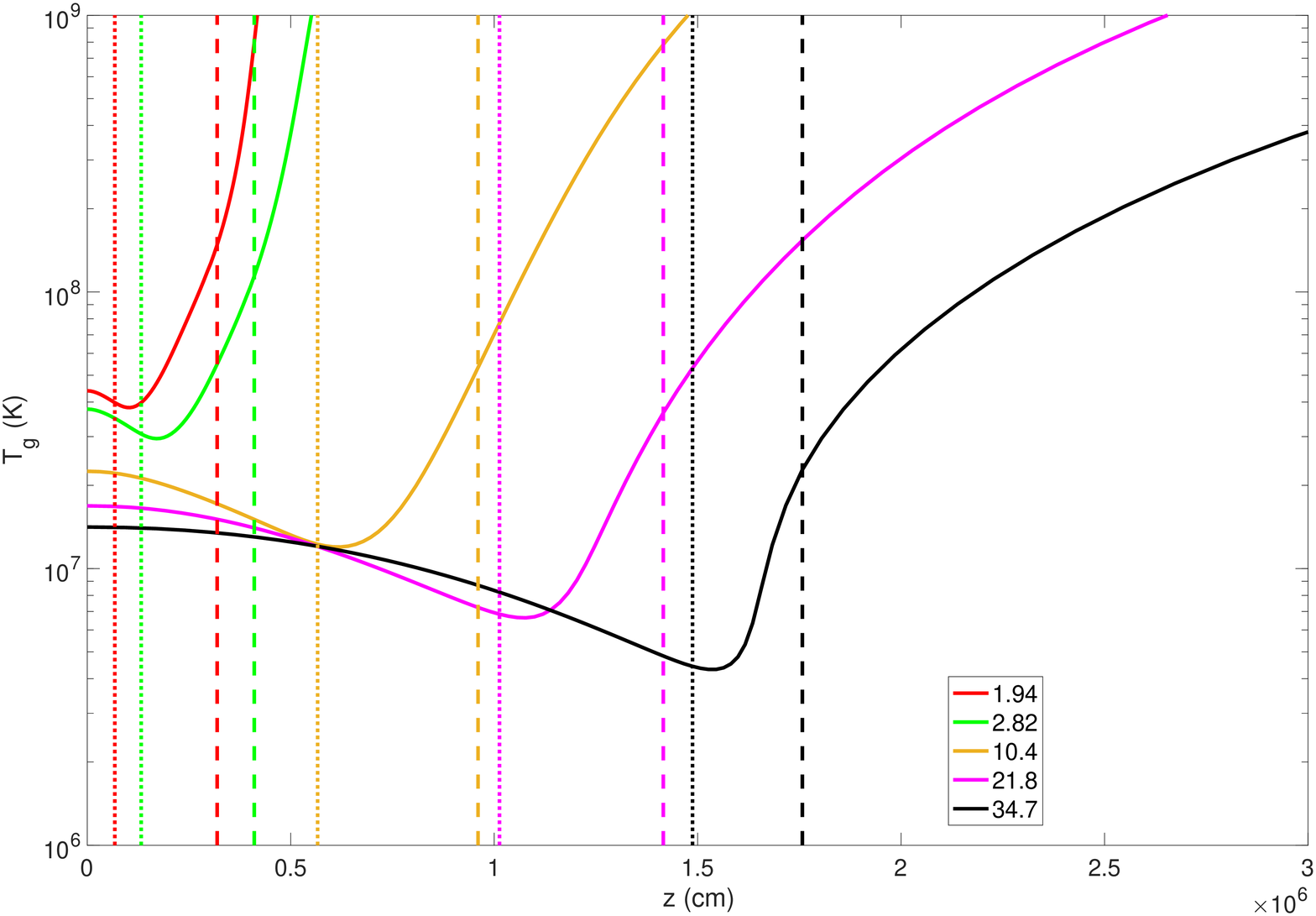}
\includegraphics[width=9.5cm]{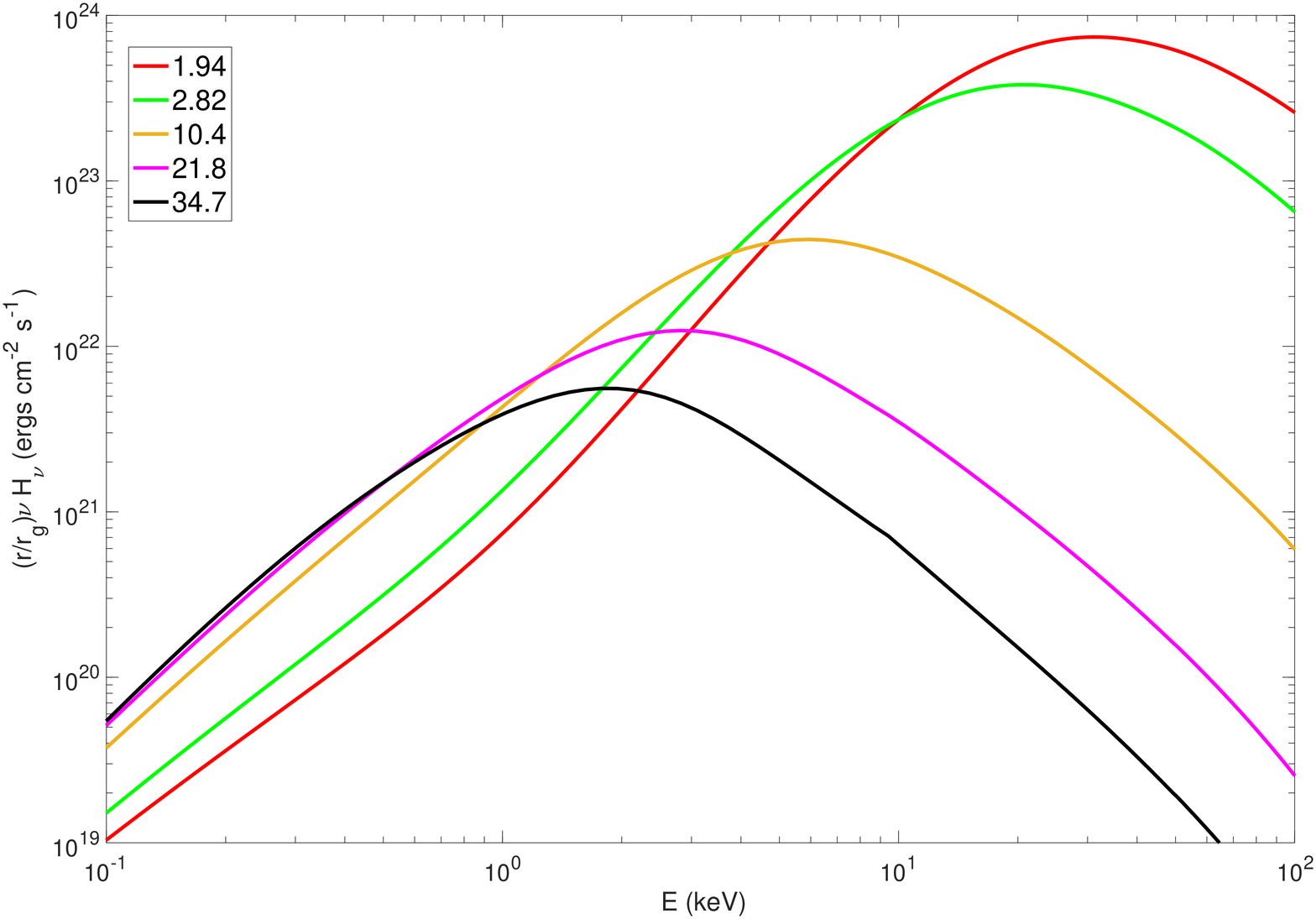}
\includegraphics[width=9.5cm]{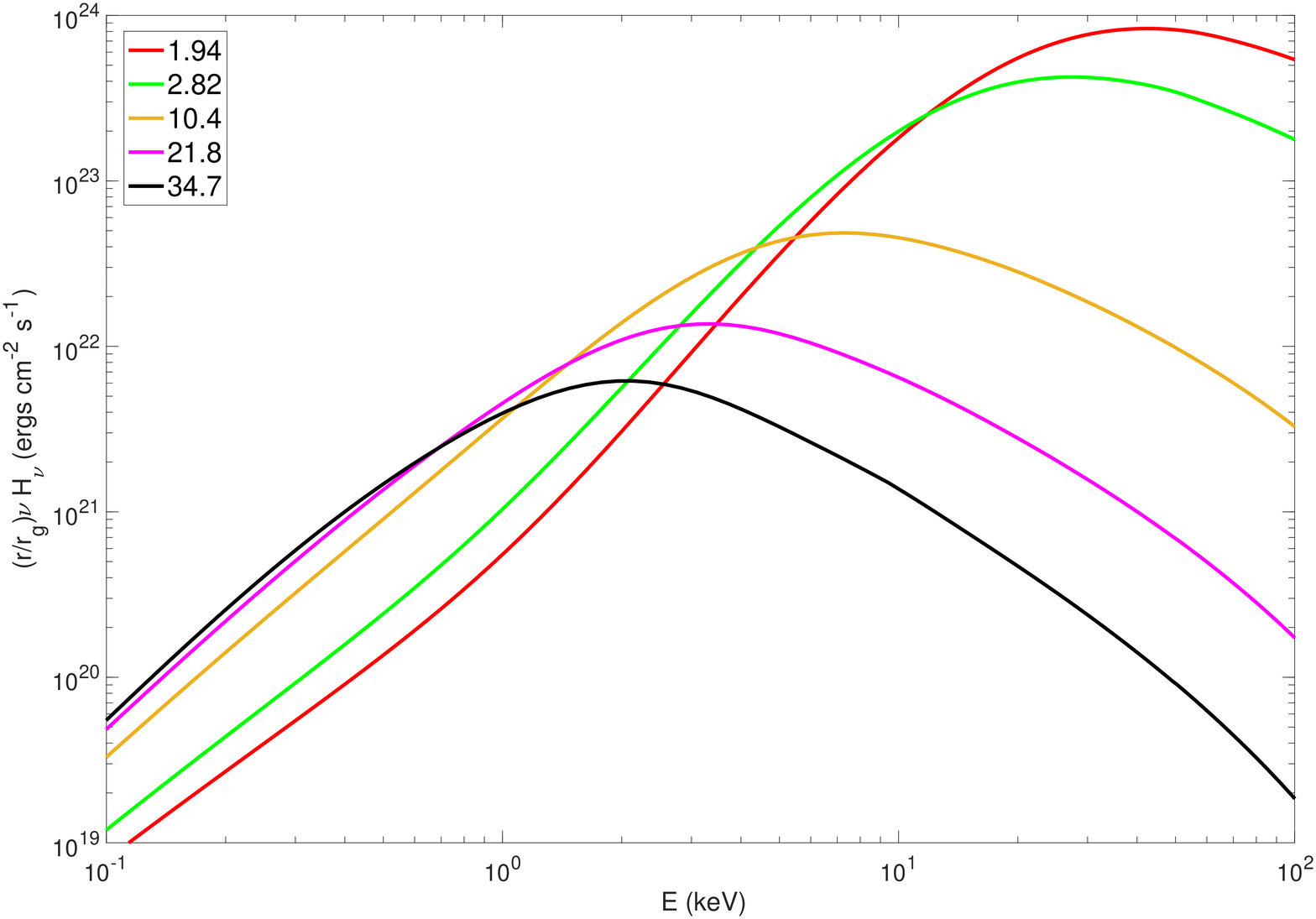}
\caption{Same as Figure (\ref{fig:a7t}) but for $a/M=0.99$.}
\label{fig:a99t}
\end{figure}

\begin{figure}
\includegraphics[width=12cm]{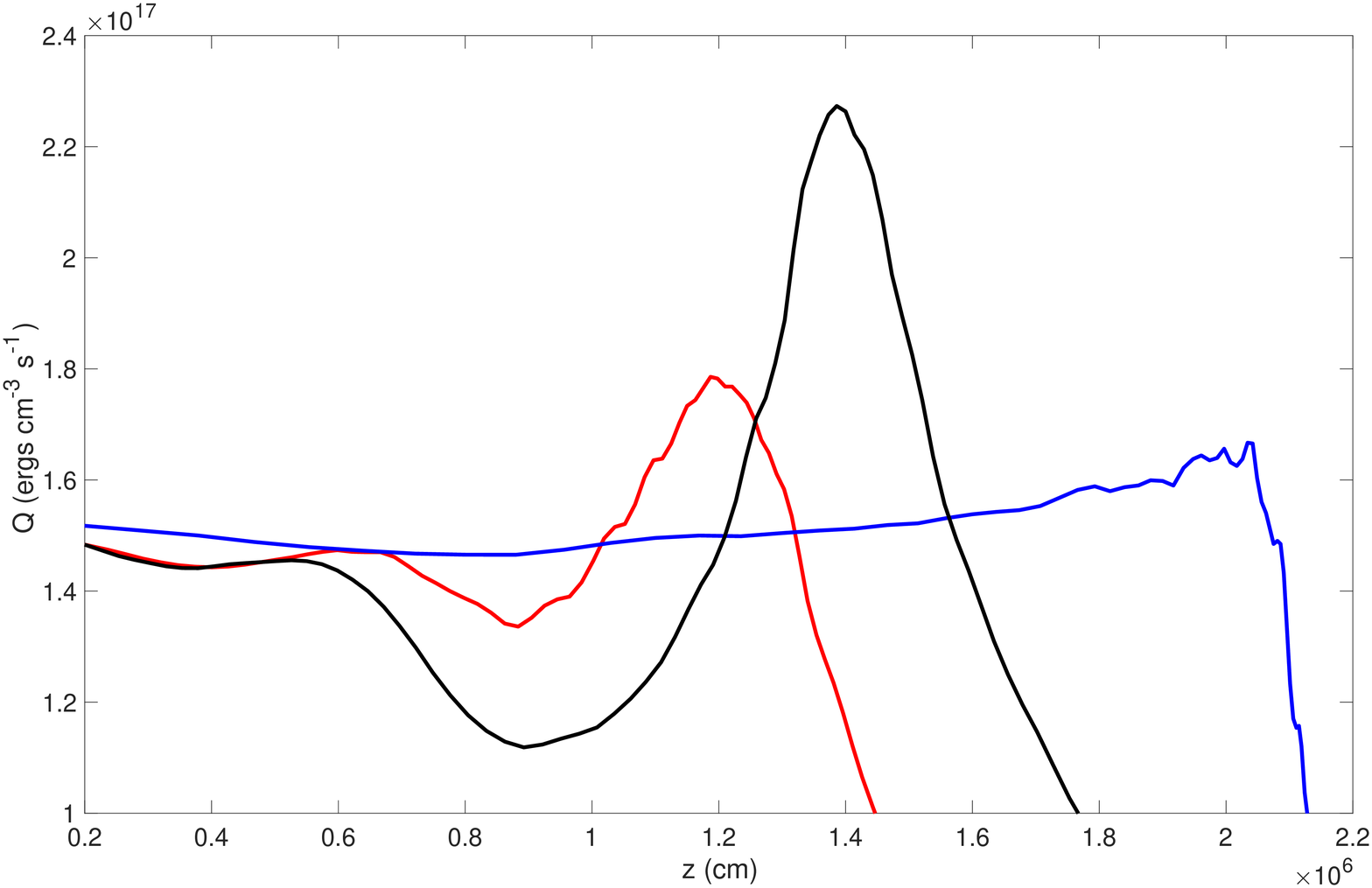}
\caption{Dissipation rate as functions of height above the mid-plane for an annulus at $r/r_g=4.13$ in a disk accreting onto an $a/M=0.7$ black hole. The black and red curves correspond to the $\xi=0.03$ and $\xi=0.1$ cases of Equation (\ref{dis2}), respectively. The blue curve is from the dissipation profile from Equation (\ref{dis1}).}
\label{fig:qz}
\end{figure}

As illustrated in Figures (\ref{fig:a7t}) and (\ref{fig:a99t}), we found that both of the single power-law dissipation prescriptions of Equation (\ref{dis2}) result in annuli spectra with high-energy tails extending to at least about $100 \ \rm keV$ regardless of black hole spin. These spectral results are consistent with the fact that these profiles put more power into the upper layers than the broken power law version described by Equation (\ref{dis1}). We further demonstrate this point in Figure (\ref{fig:qz}), which shows the locations of dissipation peaks as functions of height above disk mid-plane for all three profiles for the same annulus.

\

\begin{figure}
\includegraphics[width=9.5cm]{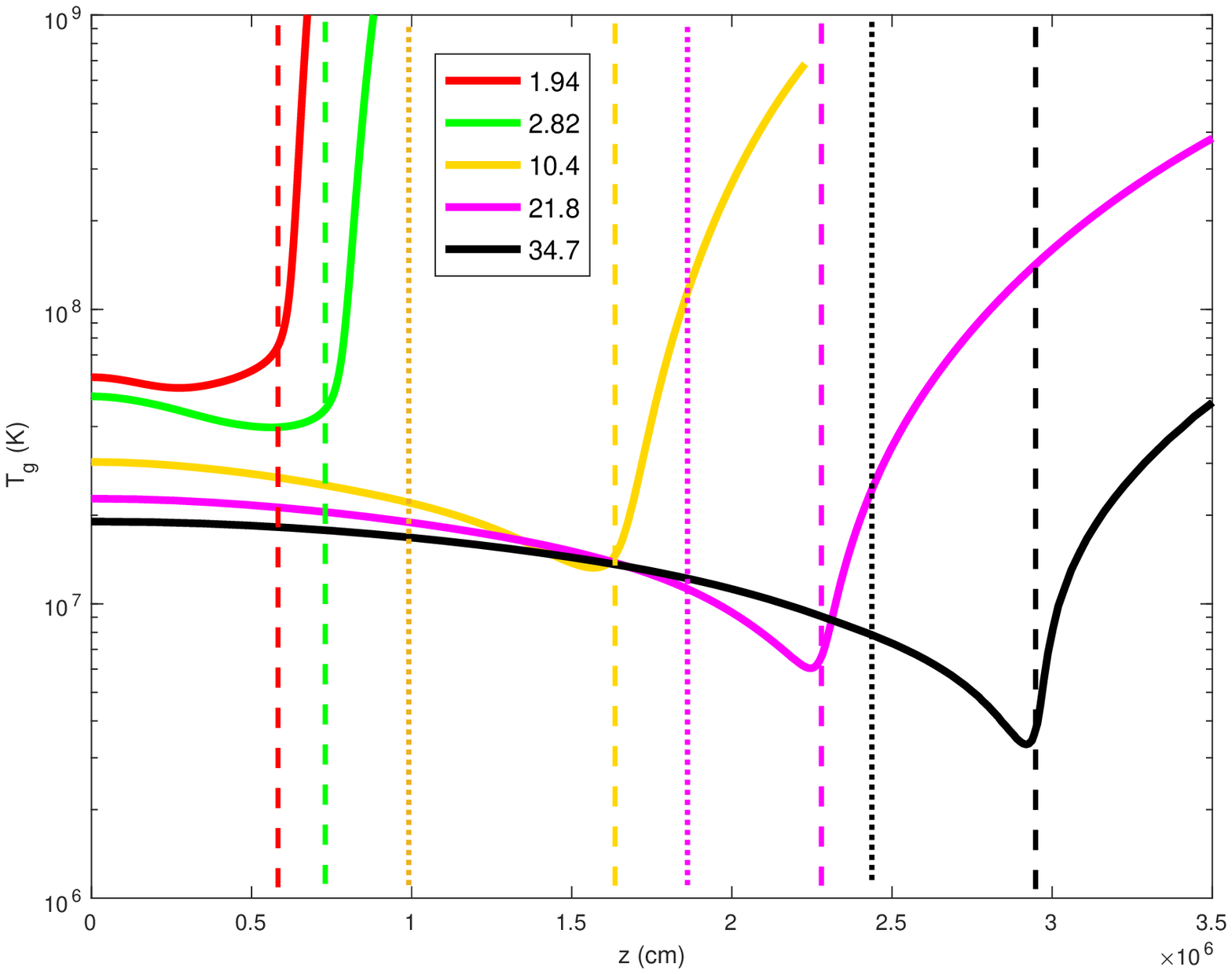}
\includegraphics[width=9.5cm]{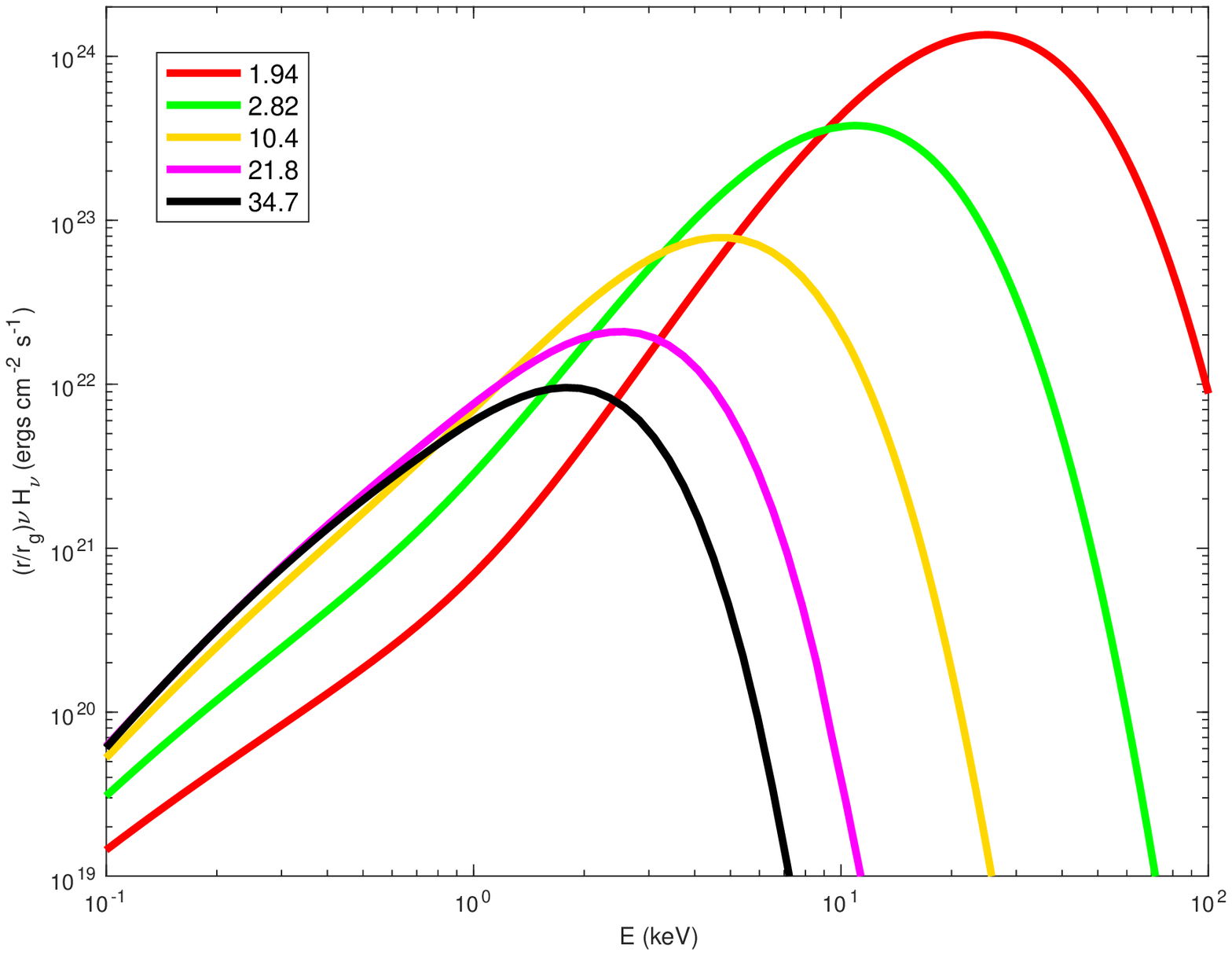}
\caption{Sample annuli temperature (left) and area-weighted spectra for disk accreting onto $a/M=0.99$ black hole with the broken-power law dissipation profile of Equation (\ref{dis1}).}
\label{fig:a99z8}
\end{figure}

The top row temperature plots in the same figures show that the gas temperature increases upwards in the annuli upper layers as the gas and radiation fall out of thermodynamic equilibrium with one another so that the gas is no longer being efficiently cooled by photons. Photons can up-scatter in this region between the temperature minimum and scattering photosphere, resulting in a non-isothermal corona over an optically thick disk. While variations of this basic geometry have been invoked to explain and model the SPL-state (for example, \cite{dk06}), merely having a upward rising temperature profile is not sufficient to produce powerful non-thermal emission. As exemplified in Figure \ref{fig:a99z8}, even with torques at the inner edge the broken power-law dissipation profile (Equation \ref{dis1}) does not lead to non-thermal spectra at any radius even for a disk accreting onto a maximally spinning black hole. 

\

\begin{figure}
\includegraphics[width=9.5cm]{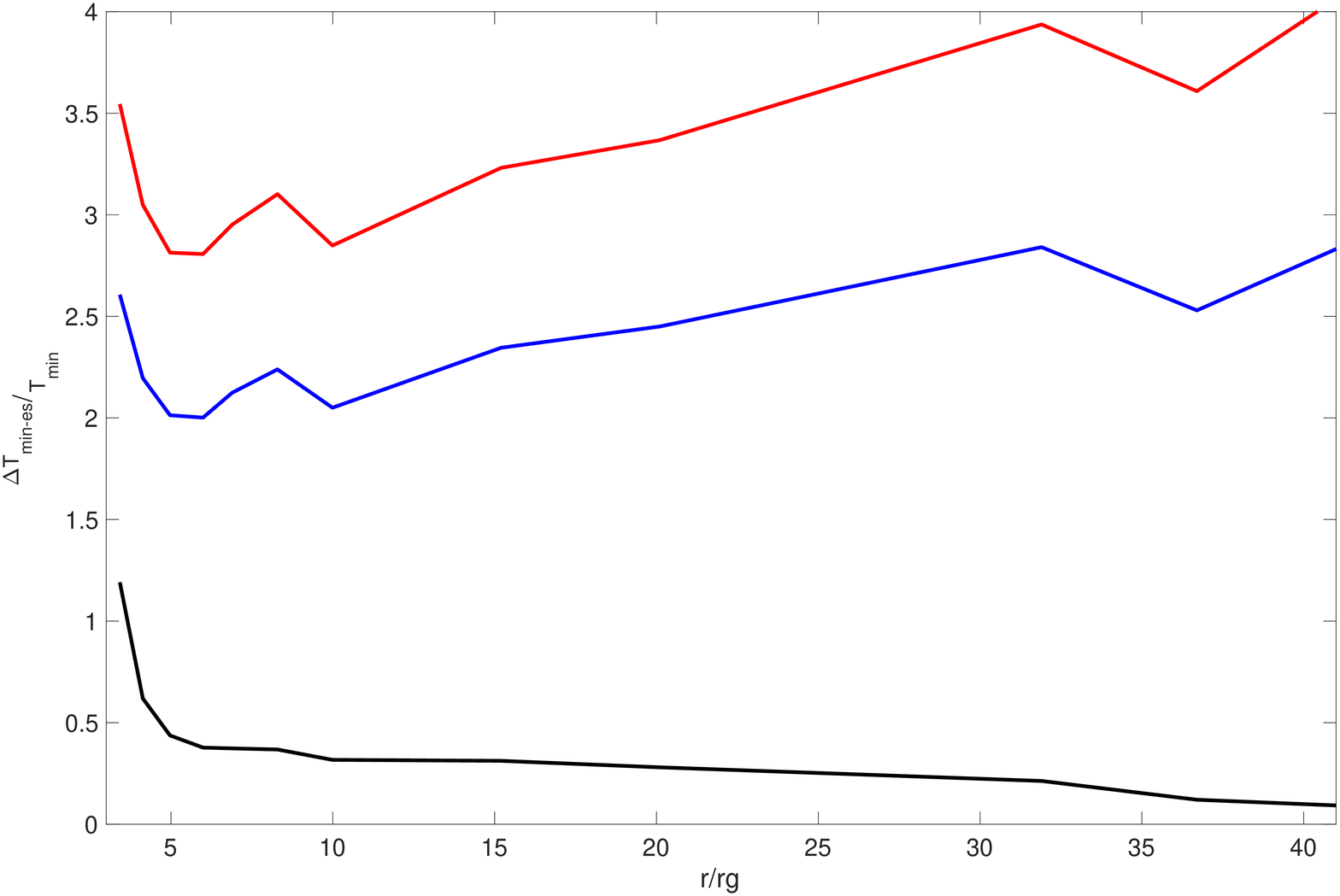}
\includegraphics[width=9.5cm]{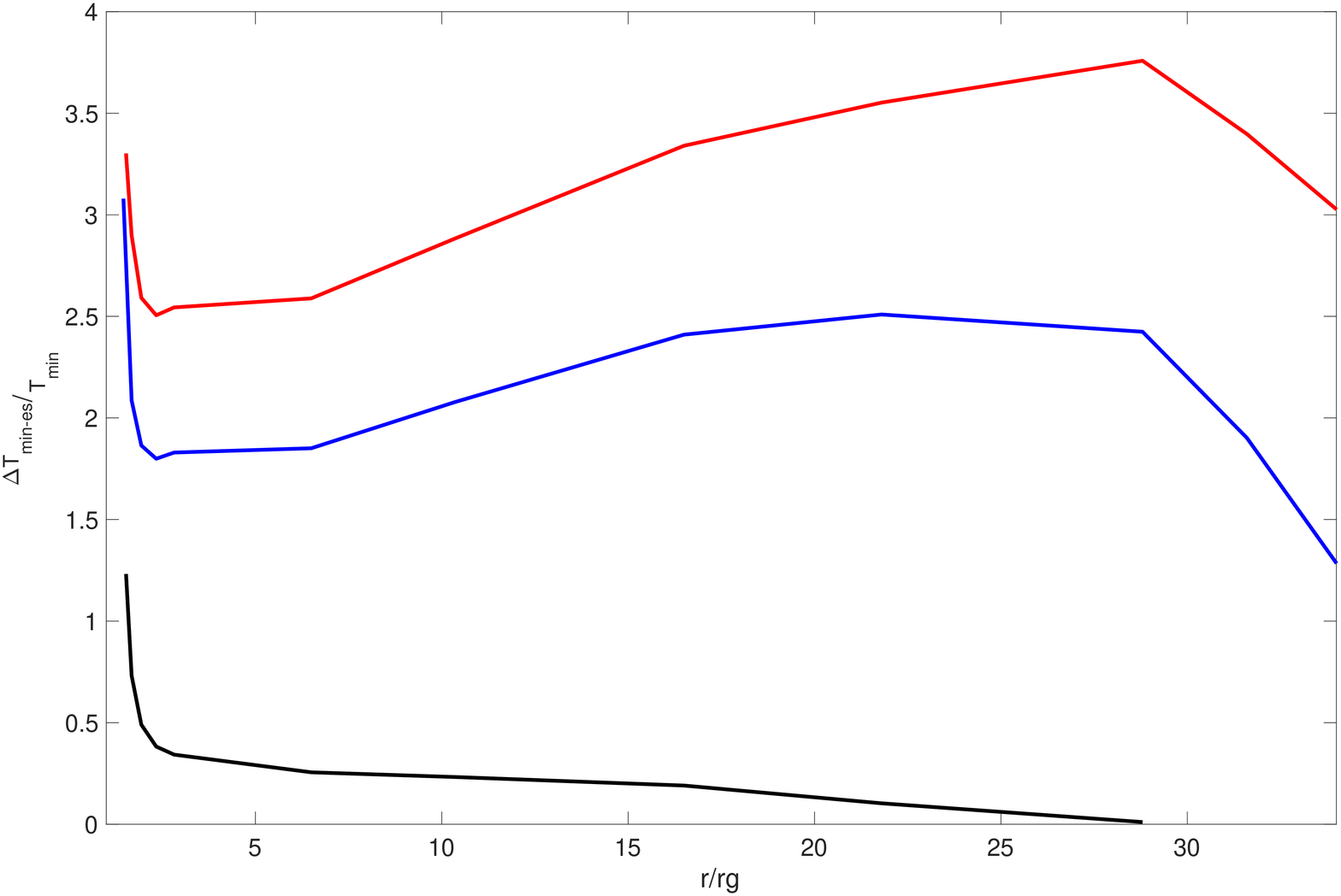}
\caption{Gas temperature $T_g(z)$ differences between the temperature minimum and scattering photosphere as functions of distance from the black hole for illustrative models with $a/M=0.7$ (left) and $0.99$ (right). The black, blue and red curves correspond to the broken power law (also used in \cite{df18}), $\zeta=0.1$ and $\zeta=0.03$ dissipation profiles, respectively.}
\label{fig:dtg}
\end{figure}

\begin{figure}
\includegraphics[width=9.5cm]{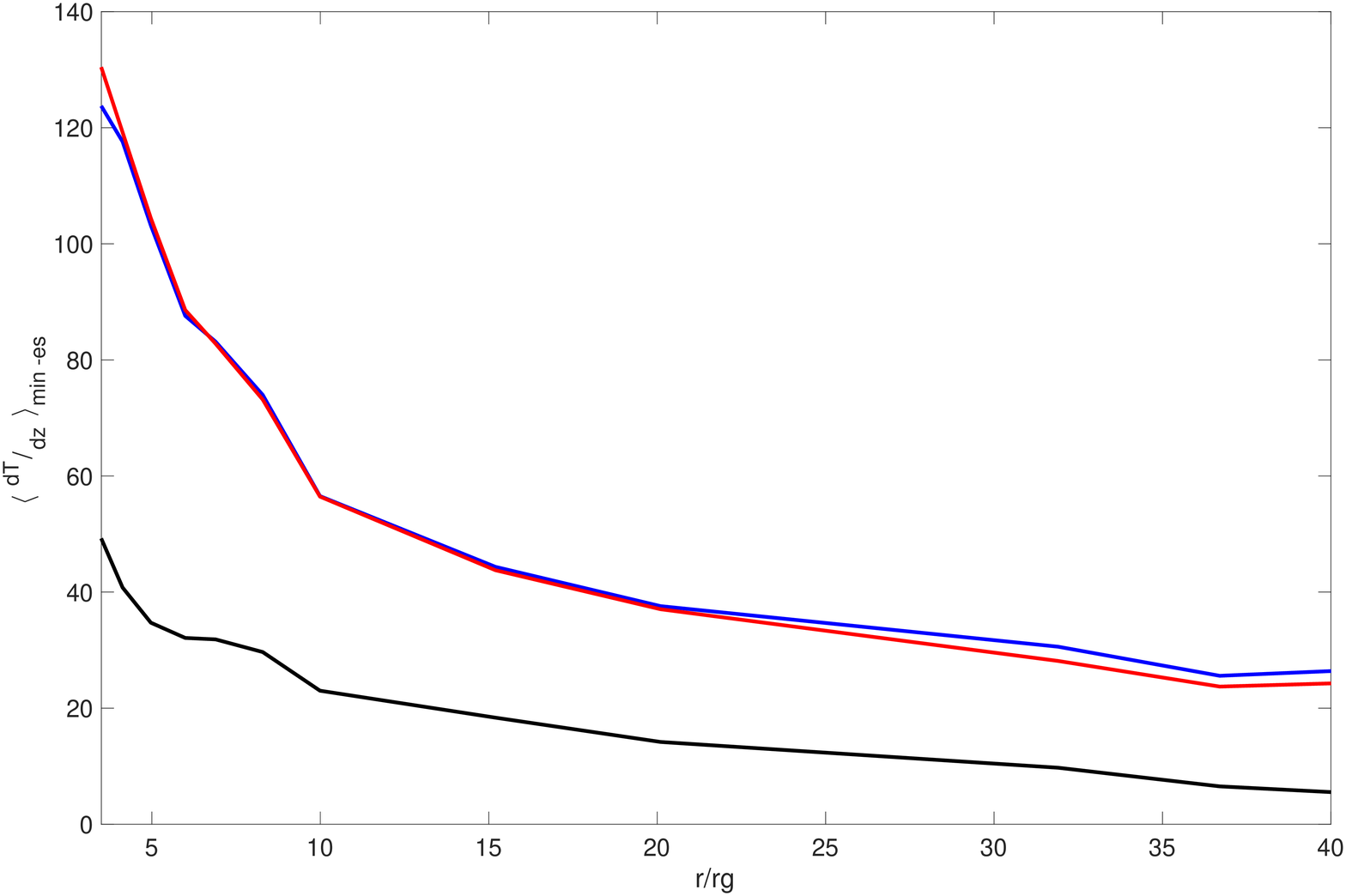}
\includegraphics[width=9.5cm]{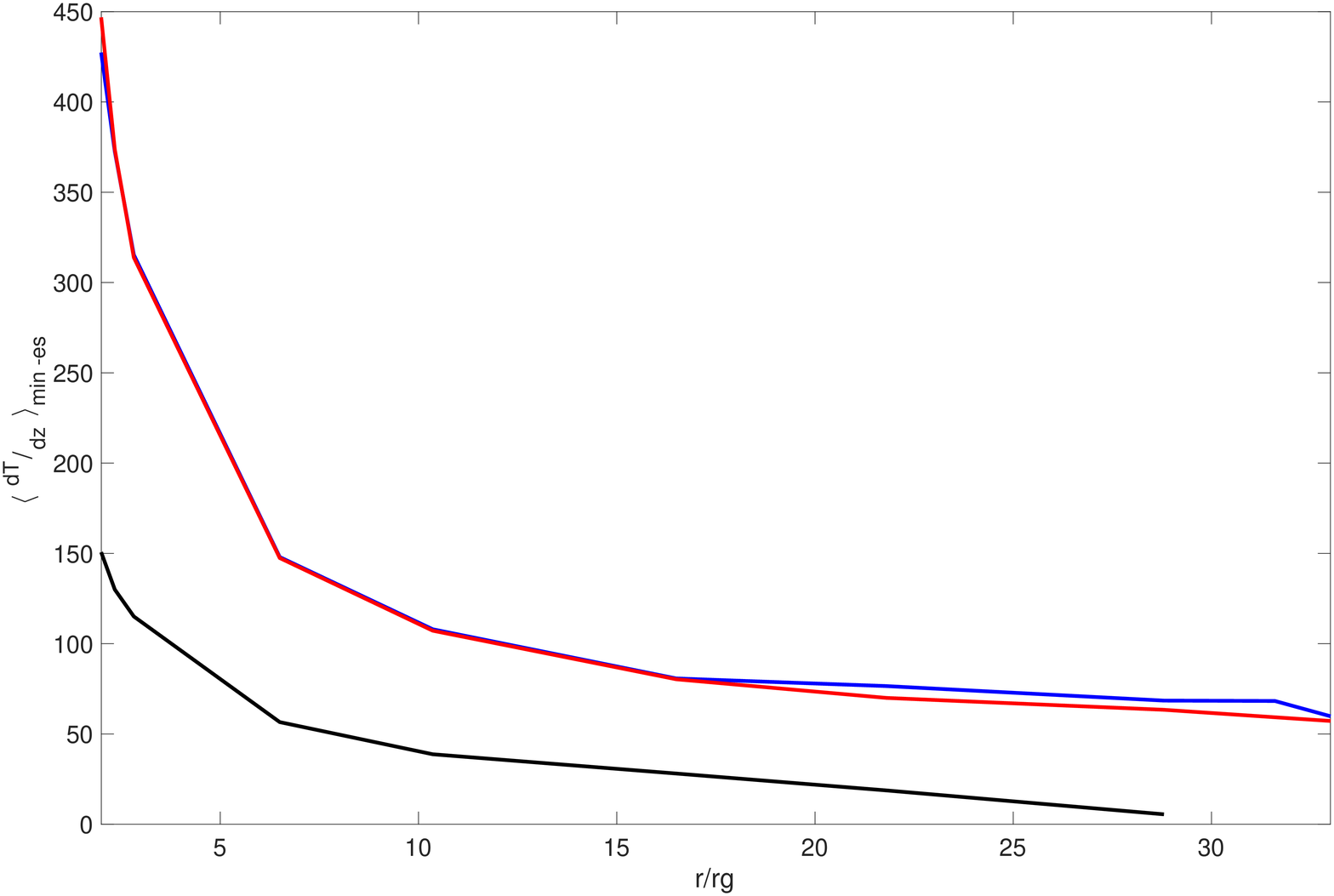}
\caption{Average slope of gas temperature $T_g(z)$ between the temperature minimum and scattering photosphere as functions of distance from the black hole for illustrative models with $a/M=0.7$ (left) and $0.99$ (right). Again, the black, blue and red curves correspond to the broken power law, $\zeta=0.1$ and $\zeta=0.03$ dissipation profiles, respectively.}
\label{fig:slope}
\end{figure}

To better understand our findings, Figure (\ref{fig:dtg}) shows that compared to the results of \cite{df18}, models with $\zeta=0.1$ and $\zeta=0.03$ all have much larger gas temperature differences $\Delta T_{\rm min-es}$ between the minima and scattering photospheres. The large temperature gap between the disk and corona coupled with increased temperatures (compared to models with less dissipation near the photospheres) at the scattering photosphere means that a Comptonized power-law tail would cover a large energy range and hence extend to higher (greater than about $100$ keV) cut-off energies. Furthermore, Figure (\ref{fig:slope}) indicates that the same additional dissipation leads to far steeper upward temperature gradients $dT_{\rm min-es}/dz$ in the photosphere. Photons therefore on average receive increasingly larger energy boosts per scattering, resulting in a harder spectrum than would be achieved within an isothermal corona. Together, these related observations imply that significant $\Delta T_{\rm min-es}$ and sharp $dT_{\rm min-es}/dz$ are both desirable for producing SPL-like non-thermal spectral tails that are simultaneously broad and hard. These results are in agreement with earlier calculations by \cite{tb13}, who performed a more limited study with only one black hole spin and did not incorporate torques at inner disk edges.

\

\begin{figure}
\includegraphics[width=9.5cm]{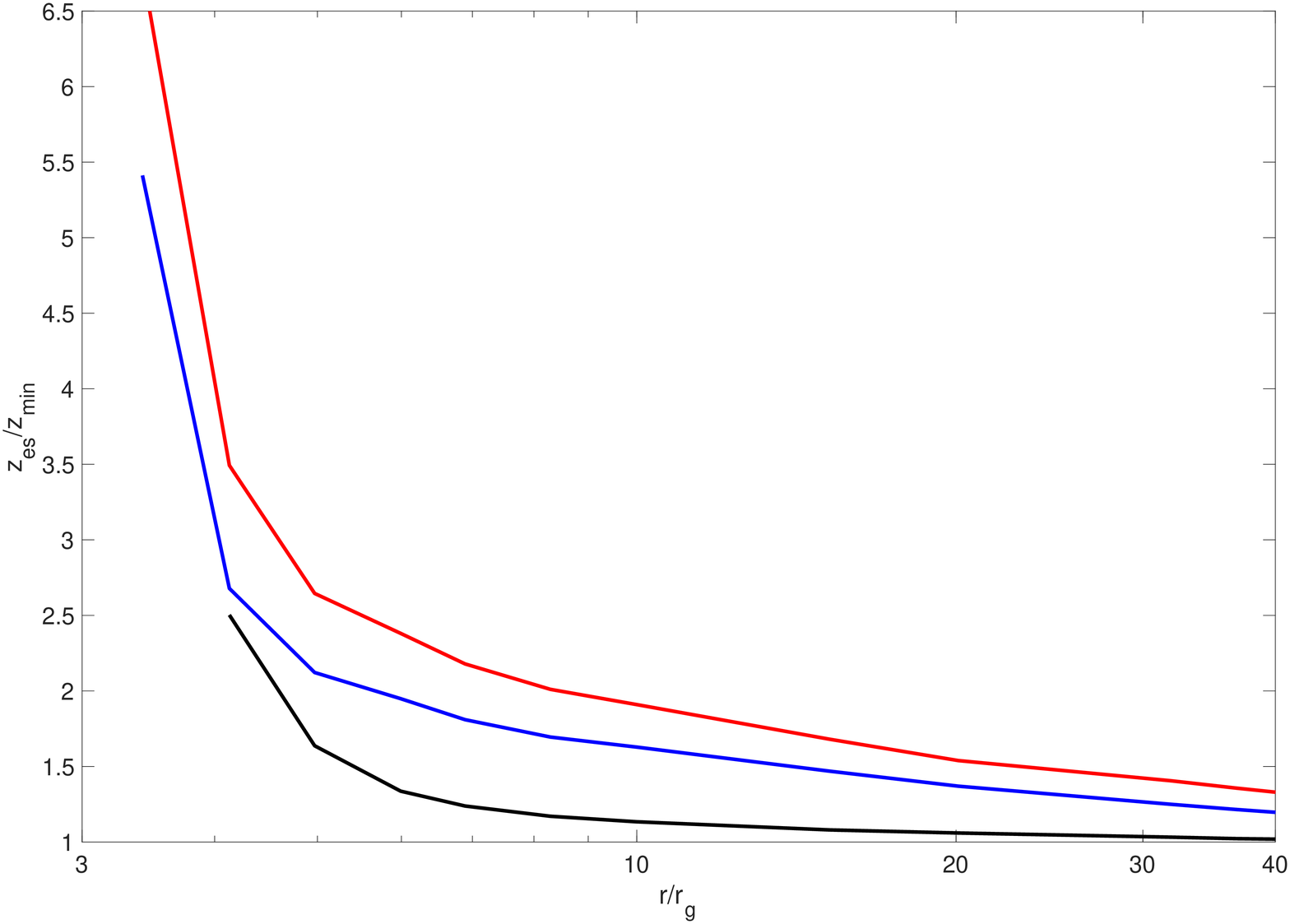}
\includegraphics[width=9.5cm]{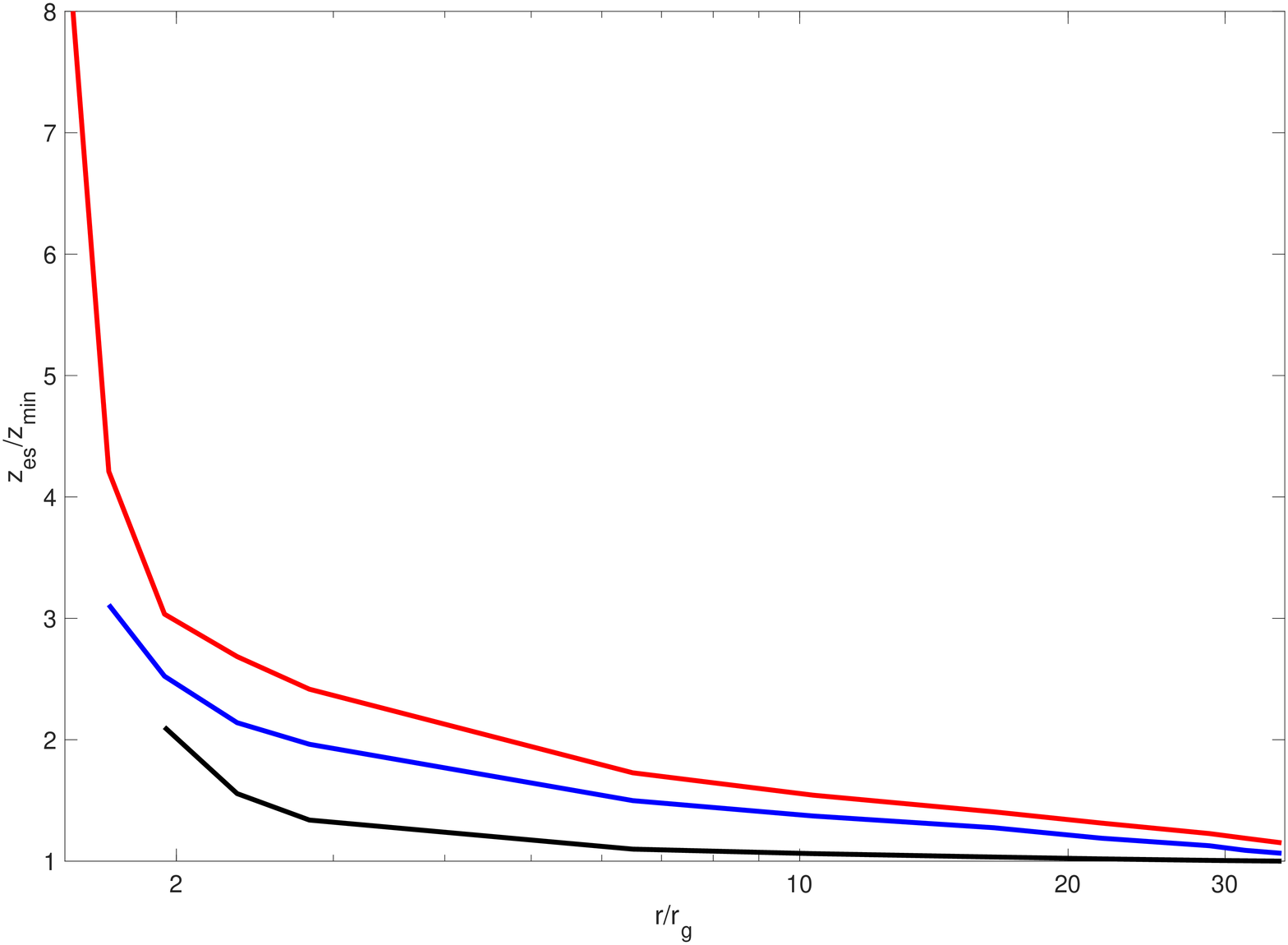}
\caption{Height of scattering photosphere $z_{\rm es}$ divided by that of temperature minimum $z_{\rm min}$ as functions of distance from black hole. The left and right panels correspond to illustrative models with $a/M=0.7$ and $0.99$, respectively. The red, blue and black colors denote $\zeta=0.03$, $\zeta=0.1$ and the broken-power-law dissipation profiles, respectively. Note that in both black curves, $z_{\rm es}/z_{\rm min}$ remains approximately $1$ for $r/r_g\agt 10$, implying that the corona is much thinner than the disk except for near the black hole.}
\label{fig:zratio}
\end{figure}

Our results are suggestive of the truncated inner disk models proposed \citep{dk06} to explain SPL observations. In such a scenario, the corona occupies an increasingly larger fraction of the total vertical height as $r/r_g$ decreases so that the spectrum from near the inner edge is dominated by Comptonised non-thermal emission. Figure (\ref{fig:zratio}) shows that the ratio of the height of the scattering photosphere to that of the temperature minimum increases sharply towards the black hole and the corona can become almost an order of magnitude thicker than the disk underneath. Moreover, $z_{\rm es}/z_{\rm min}$ reaches larger maximum value when the fraction of accretion power dissipated away from the mid-plane increases (lower $\zeta$). When $\zeta=0.1$ and $0.03$, the corona is geometrically significant for a much larger portion of the disk than models with the broken power-law dissipation profile. 

\

\subsection{Disk Integrated Spectra}

\begin{figure}
\includegraphics[width=9.5cm]{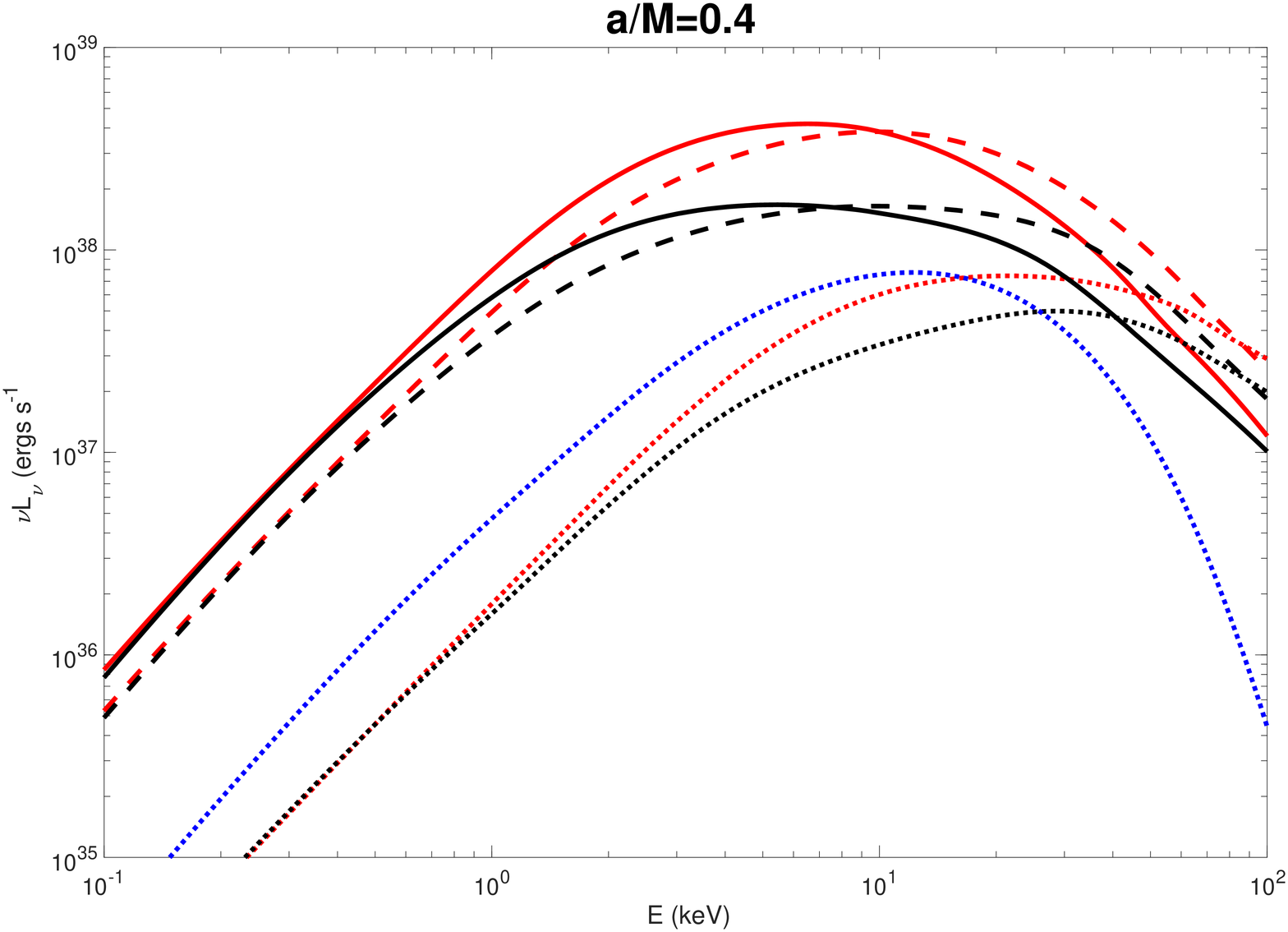}
\includegraphics[width=9.5cm]{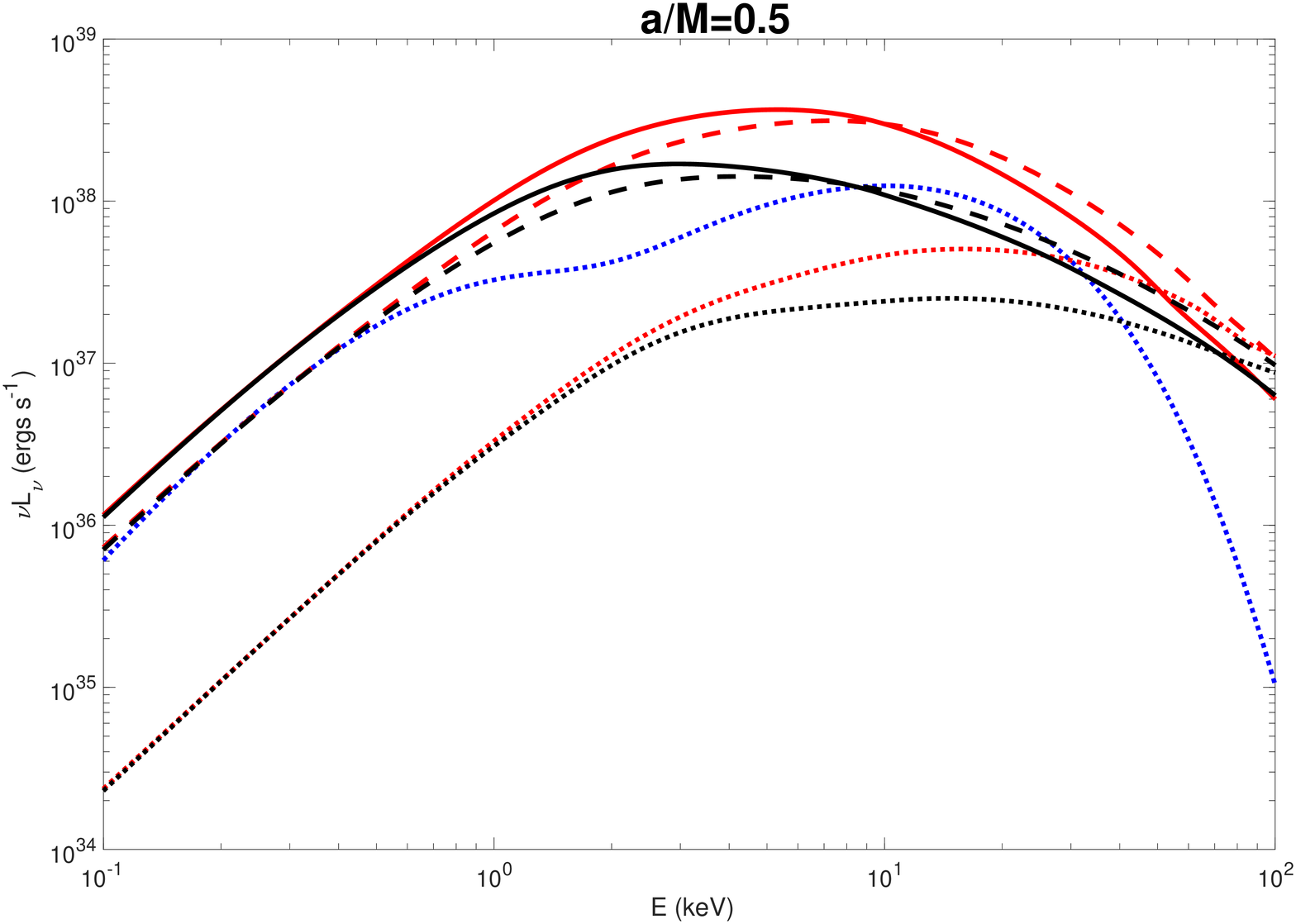}
\includegraphics[width=9.5cm]{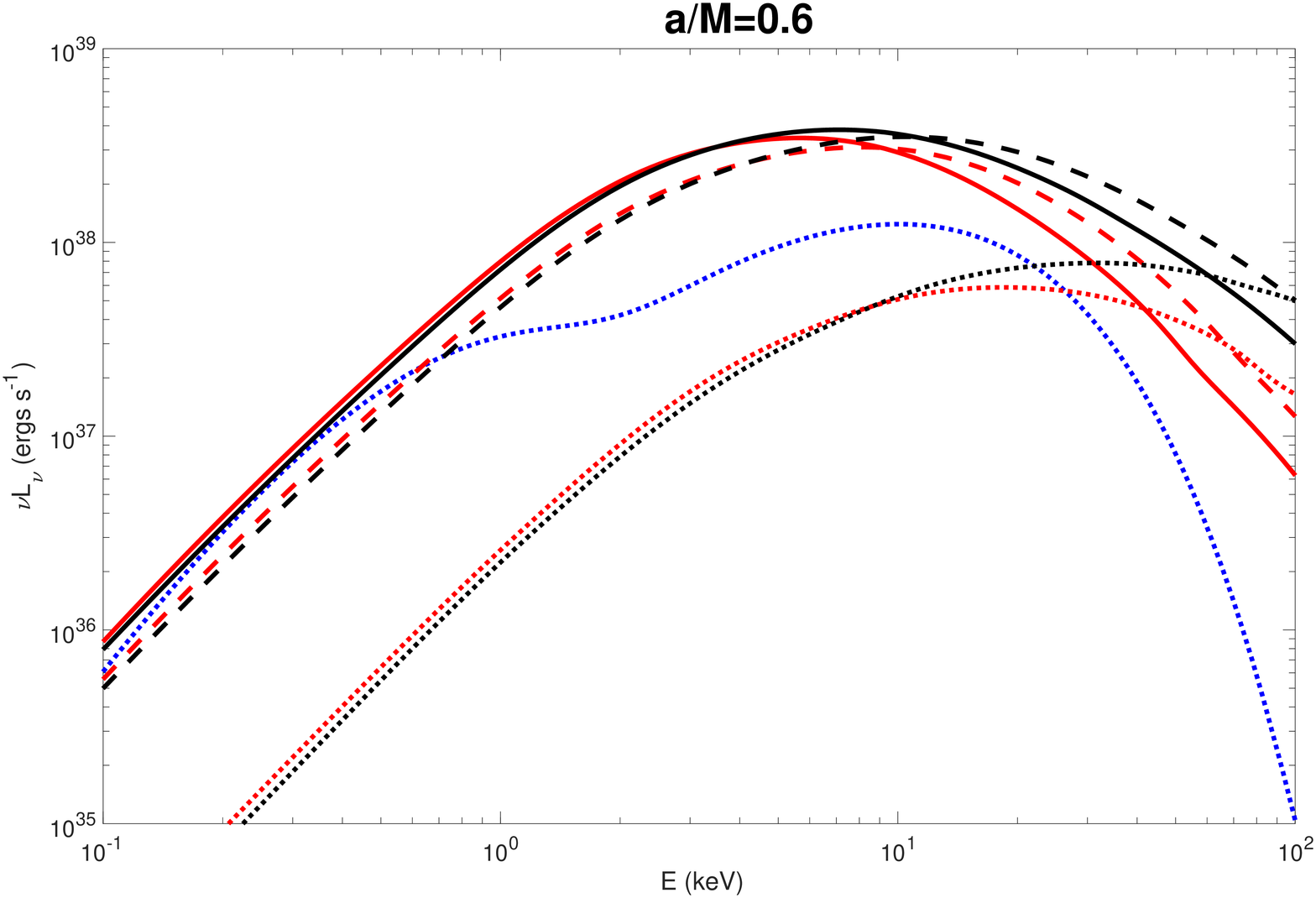}
\includegraphics[width=9.5cm]{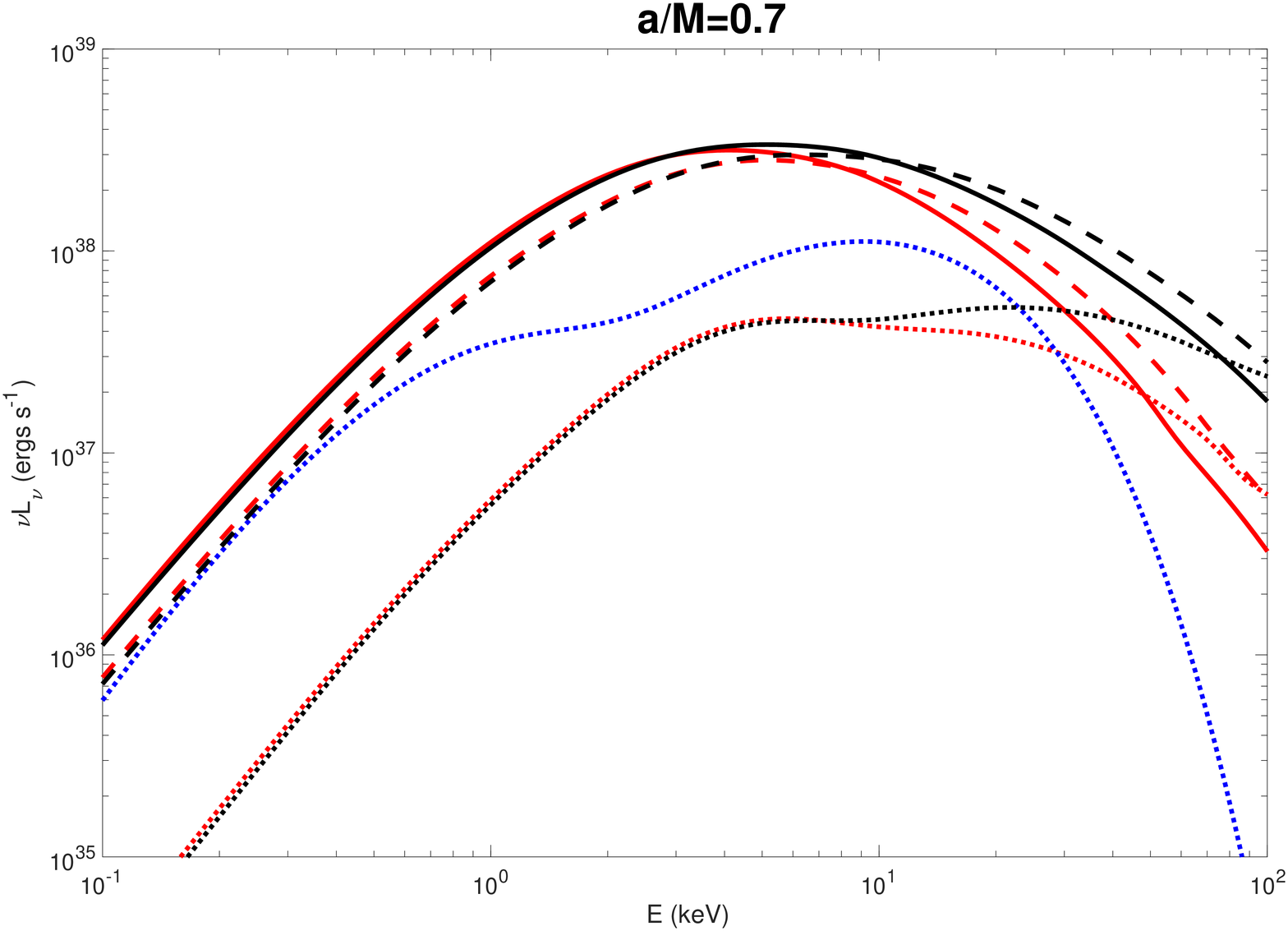}
\includegraphics[width=9.5cm]{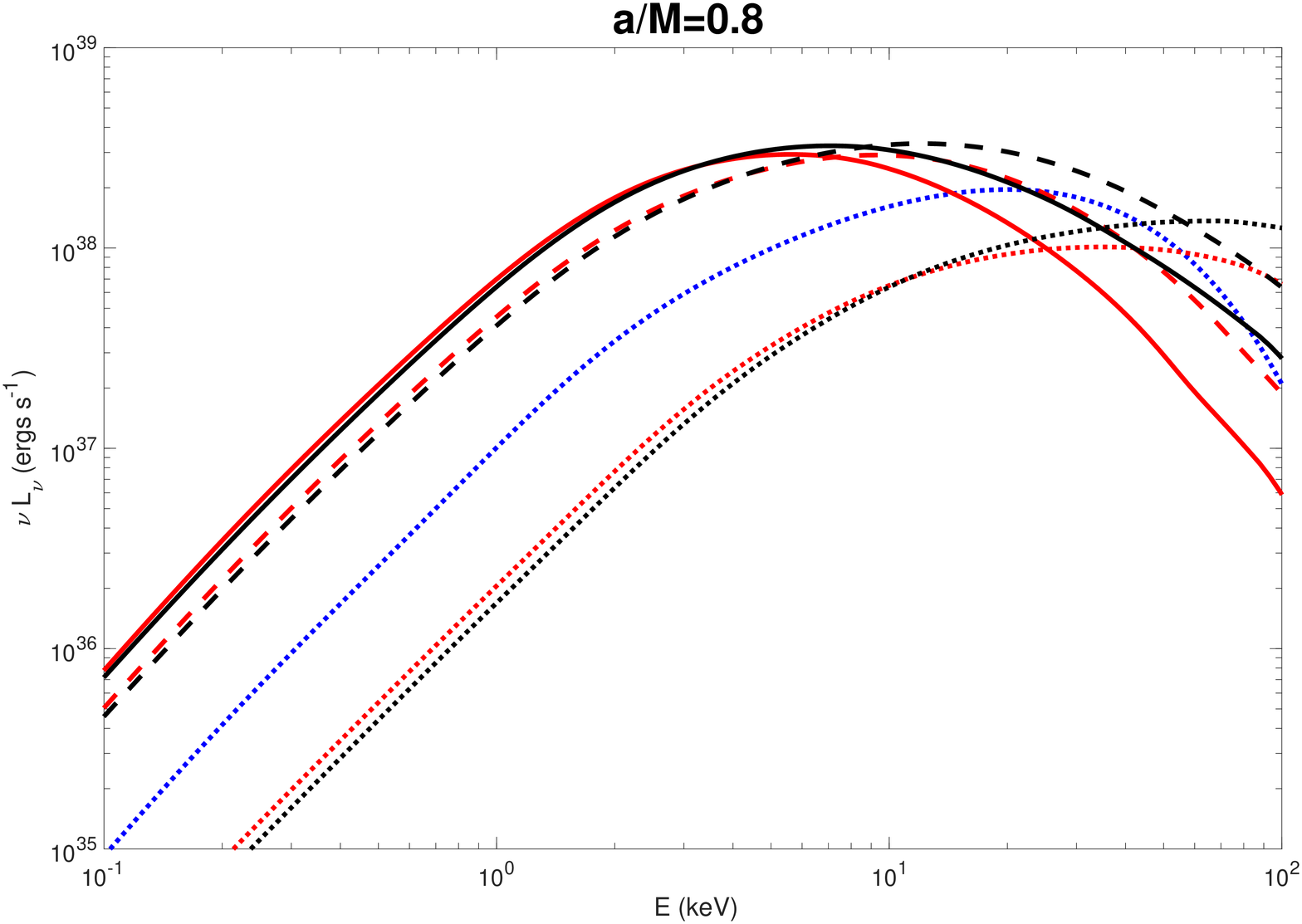}
\includegraphics[width=9.5cm]{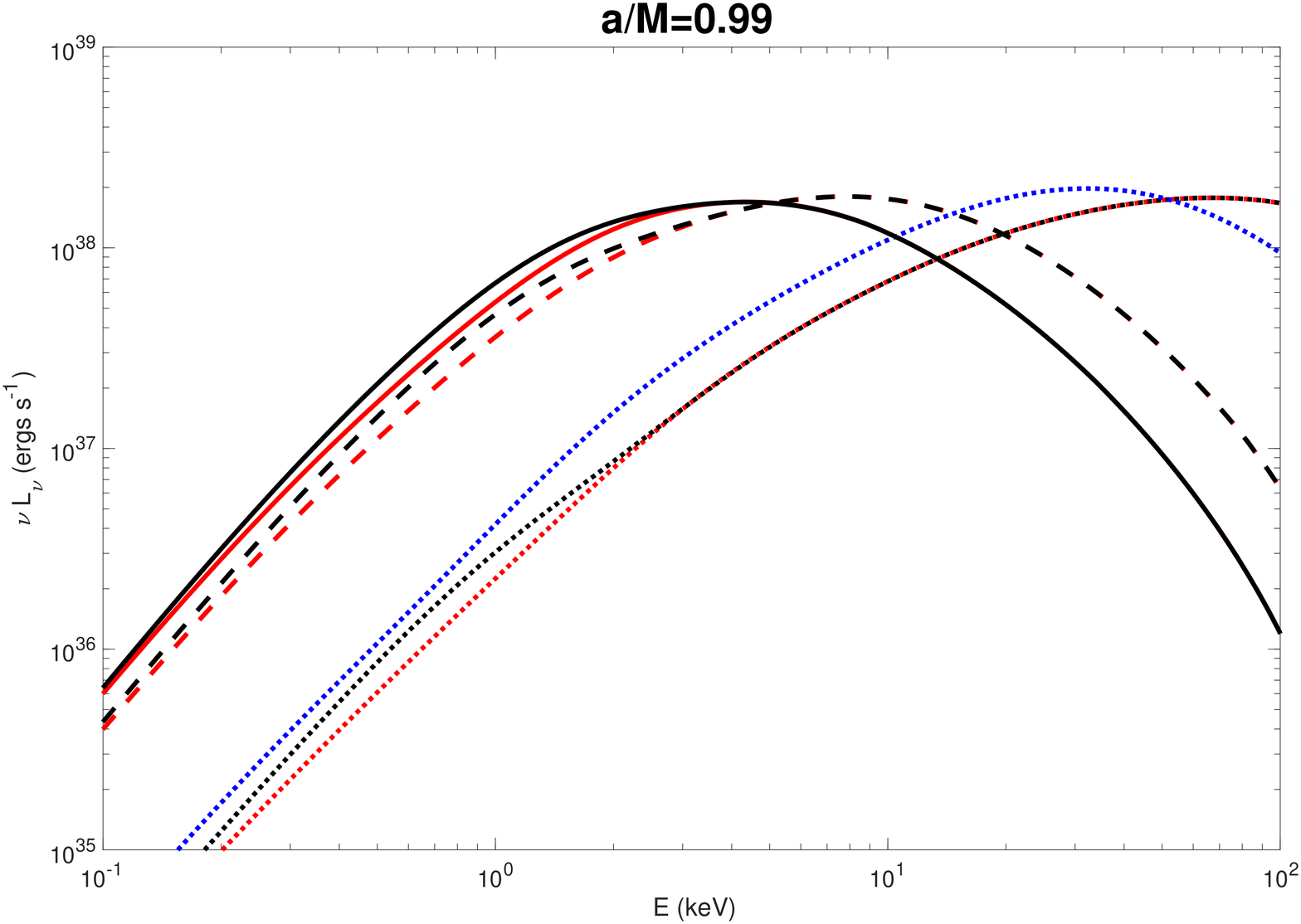}
\caption{Full-disk spectra as seen by distant observers for accretion onto black holes with various spins. The red and black curves denote dissipation profiles with $\zeta=0.1$ and $0.03$, respectively. The dotted, dashed and solid lines correspond to viewing the disk nearly edge on, at inclination angle $\phi=\pi/4$ (measured with respect to the line of sight of the observer) and face-on, respectively. For comparison, we also over-plotted edge-on spectra for disks with the broken-power law dissipation prescription (dotted blue).}
\label{fig:fd}
\end{figure}

We present full-disk spectra for accretion with six different black hole spins in Figure (\ref{fig:fd}). In all cases, the more aggressive $\zeta=0.1$ and $0.03$ dissipation prescriptions result in clearly non-thermal spectra with energetically significant tails extending to at least $100 \ \rm keV$ regardless of inclination. On the other hand, models with the broken power-law dissipation profile (blue curves) generally do not display powerful spectral tails even when viewed edge-on. This is most likely because incorporating stresses at the inner disk edge alone does not spread the annuli spectral peaks over a sufficiently large energy range to produce a significant power-law tail \citep{df18}. The only exception is the $a/M=0.99$ disk due to strong relativistic effects. More generally, in edge-on disks the same relativistic effects lead to energetic tails with photon index $\Gamma-2$ (defined such that $\nu L_{\nu}\propto\nu^{-(\Gamma-2)}$) that approximately agrees with SPL data. However, the full-disk spectra from these models peak at much higher energies ($\ga 20 \ \rm keV$) compared to observations ($\la10 \ \rm keV$).

\

On the other hand, at more moderate inclinations (face-on and at $45$ degrees) disks with the most aggressive $\zeta=0.03$ dissipation profile do have both peak energies and photon indices that broadly agree with some SPL observations. Moreover, such models are not isolated occurrences and exist for black hole spins ranging from $a/M=0.5$ to $0.8$. For example, the face-on $a/M=0.6$ spectrum has $\Gamma\approx2.7$ and $E_{\rm peak}\approx 7 \ \rm keV$, which are close to measurements from the SPL outburst of GRO J1655–40 \citep{mr06}.

\

\begin{figure}
\includegraphics[width=14cm]{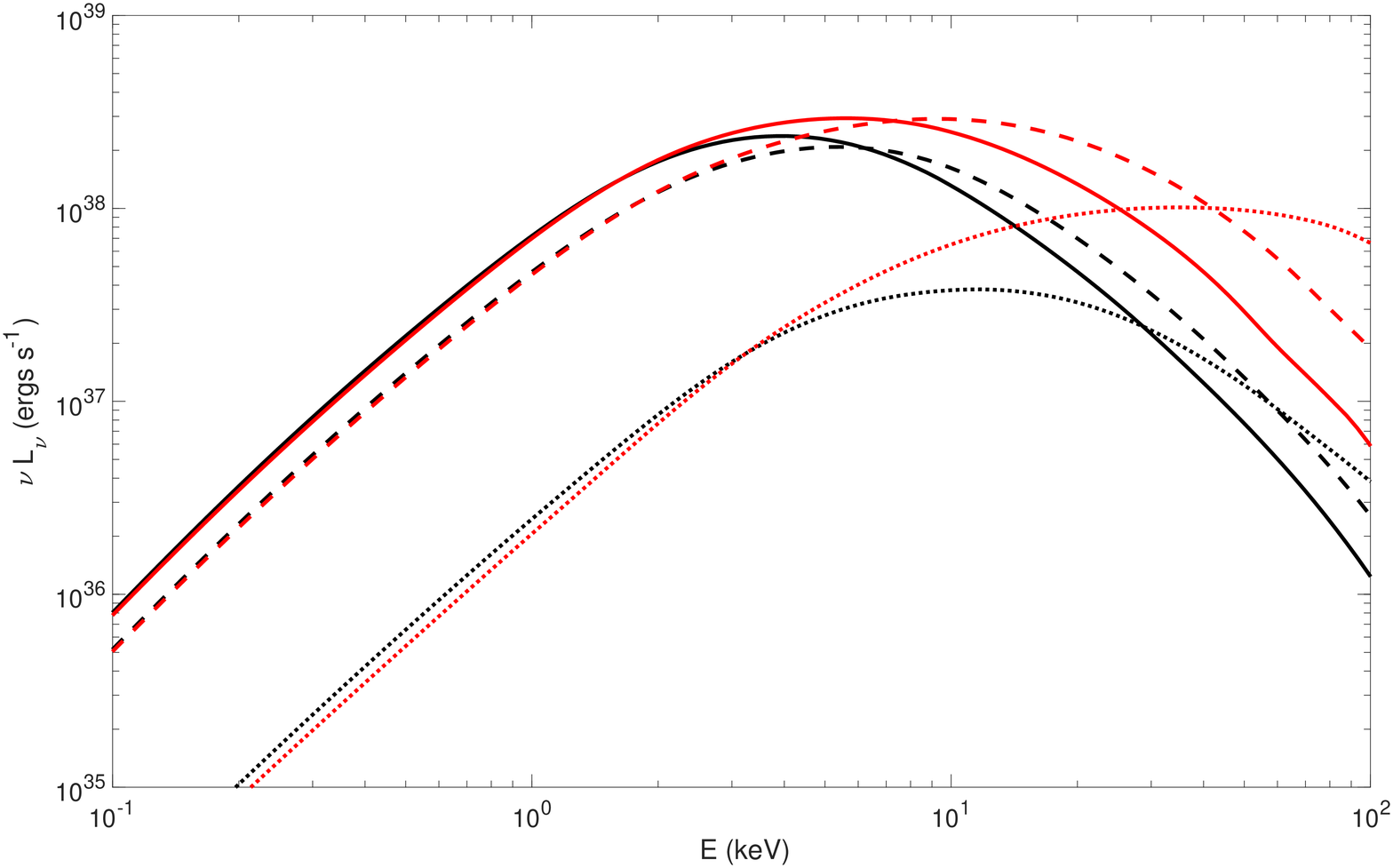}
\caption{Full-disk spectra for a $\xi=0.1$ disk around an $a/M=0.8$ black hole with (red) and without (black) inner torque. The solid, dashed and dotted lines indicate face-on, $\phi=\pi/4$ and edge-on viewing, respectively.}
\label{fig:fdcompare}
\end{figure}

For comparison, we note that models with increased photospheric dissipation but zero inner torque have spectral peaks at lower photon energies but still exhibit powerful non-thermal tails. Stresses at the ISCO causes annuli effective temperature to rise sharply towards the black hole, which as illustrated in Figure (\ref{fig:fdcompare}) in turn results in full-disk spectra peaking at higher frequencies.

\

Despite these promising results, we stress that thermal Comptonisation is unlikely able to explain all SPL observations. While numerical considerations limited our upper photon energy to $\approx 100 \ \rm keV$, we expect most of our face-on spectra (except models with maximally spinning black holes) to exhibit a spectral cut-off at higher energies that correspond to the scattering photosphere temperatures of the inner most annuli. However, some BHB systems show unbroken power-law tails extending to hundreds of keVs or even MeVs \citep{gr98, zd01, lw05}. It is unlikely that relativistic effects alone can produce such energetic emission when the corona temperature tops out at about $100 \ \rm keV$, regardless of viewing angle and black hole spin.

\subsection{Relativistic Effects}

\begin{figure}
\includegraphics[width=14cm]{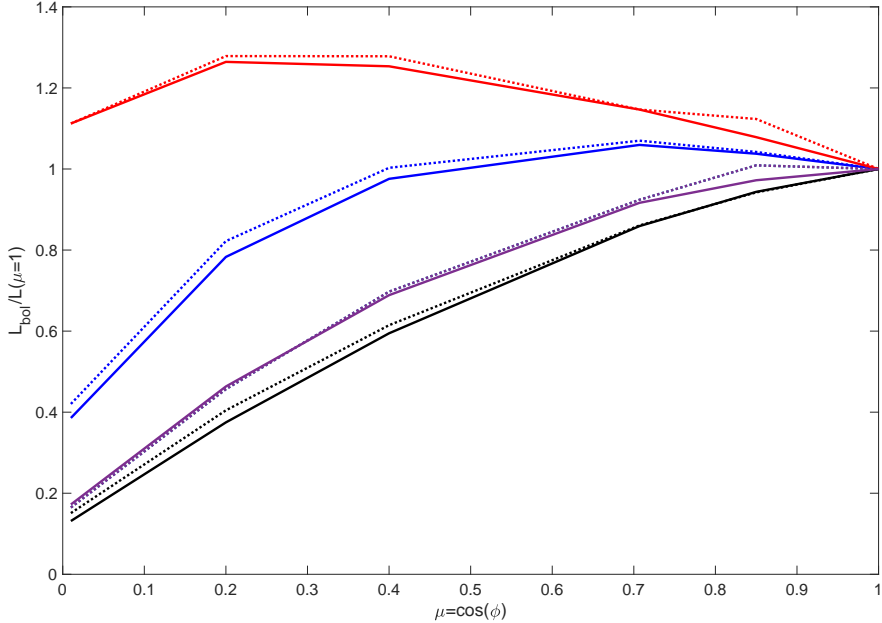}
\caption{Normalized bolometric luminosity as functions of cosine of viewing angle $\phi$, defined such that $\mu=0$ is edge-on while $\mu=1$ is face-on. The solid and dotted lines denote models using the $\zeta=0.1$ and $0.03$ dissipation profiles, respectively. The different groups of curves correspond to accretion onto black holes with $a/M=0.4$ (black), $0.7$ (purple), $0.8$ (blue) and $0.99$ (red).}
\label{fig:limb}
\end{figure}

As pointed out by \cite{c75} and \cite{ak00}, relativistic beaming and bending enhances radiation from the disk equatorial plane, which competes with the limb darkening effect that is proportional to $\mu=\cos(\phi)$, where $\phi$ is defined such that $\mu=1$ correspond to face-on viewing. As shown in Figure (\ref{fig:limb}), this focusing is more pronounced at higher black hole spins but only depends weakly on choice of dissipation profiles. At $a/M=0.99$, relativistic effects are so strong that the bolometric luminosity actually rises as inclination increases until the reduced effective emission area finally becomes a significant influence at nearly edge-on. Our findings qualitatively agree with those of \cite{ak00}, who found that the disk may actually become limb brightened as $\mu$ approaches $0$ when the radiative efficiency enhancement $\D\ep\ga 1$. Finally, we also observe the progressively stronger spectral hardening in the edge-on spectra with rising $a/M$ as predicted by \cite{c75}, again largely independent of dissipation prescriptions.

\subsection{Quasi-periodic Oscillations}

\begin{figure}
\includegraphics[width=14cm]{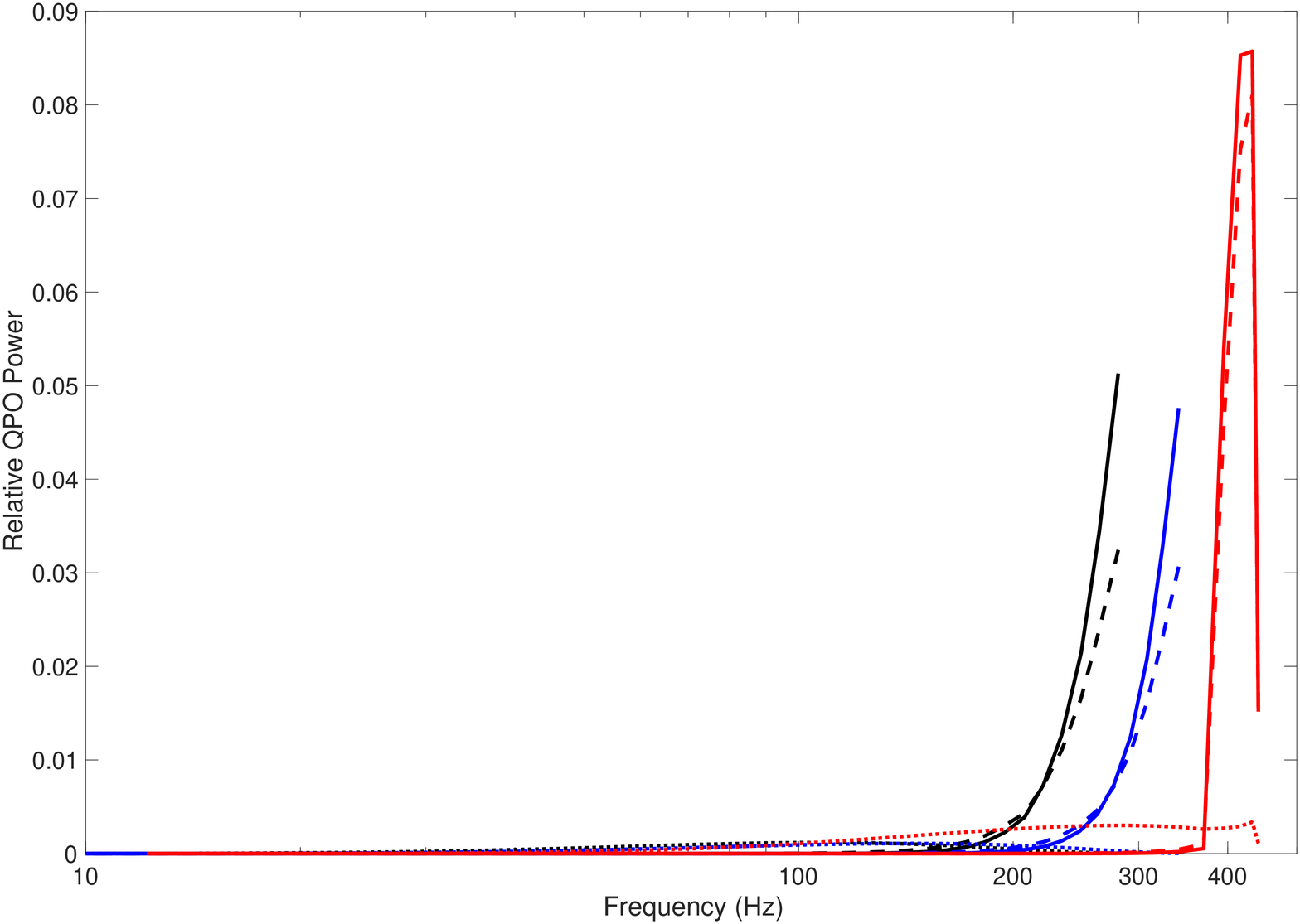}
\caption{Sample synthetic QPO power spectra for models with $\zeta=0.1$ integrated over several photon energy bands as functions of vertical epicyclic frequency. The black, blue and red curves denote accretion onto black holes with $a/M=0.5$, $0.7$ and $0.99$, respectively. The solid, dashed and dotted lines correspond to integration over $13 - 30 \ \rm keV$, $6 - 30 \ \rm keV$ and $0.1 - 2 \ \rm keV$ bands, respectively.}
\label{fig:qpo}
\end{figure}

\cite{db14} proposed that accretion flows with non-zero stresses at the ISCO can naturally gave rise to HFQPOs without the need for a large portion of the disks to oscillate coherently. To obtain synthetic QPO power spectra, we assume that each disk annulus oscillate at its respective vertical epicyclic frequency whose radial dependence is determined only by the spacetime geometry near the central object. We then compute the the relative QPO profile (Figure \ref{fig:qpo}) by calculating the flux from an individual annulus in a particular photon energy band normalized to the total flux of the disk radiated in that band, and finally squaring to convert to power.  With the additional accretion power dissipated in the annuli upper layers, the quality factors of the $13 - 30\ \rm keV$ band range from $3$ to $10$, which are higher than those found by \cite{df18}. More significantly, we obtained high quality factor QPOs in the same non-maximally spinning models whose spectral indices are consistent with SPL data, which suggests that both non-zero inner torque and additional dissipation near annuli photospheres are necessary to explain observations. 

\section{Discussions and Conclusions}

We computed vertical structures and emergent spectra of annuli models appropriate for near-Eddington accretion flows onto stellar mass black holes. Our approach self-consistently couples the radiative transfer and vertical structure equations at each radial annuli grid point without assuming color corrections to account for deviations from blackbody spectra. We greatly expanded parameter range coverage compared to previous studies. We demonstrated that incorporating both magnetic stresses at the ISCO and increased dissipation in disk upper-layers could result in models that simultaneously exhibit SPL-like spectra and HFQPOs, in agreement with observations.

\

We found that models with moderate black hole spins ($a/M<0.99$) and highest dissipation ($\zeta=0.03$) in the disk upper-layers can potentially produce SPL-like spectra in terms of both peak energy and power-law index, provided the viewing inclination is sufficiently moderate that relativistic effects do not push the peak to beyond $10 \ \rm keV$. Specifically, the dissipation profiles with $\zeta=0.1$ and $0.03$ all result in clearly non-thermal spectral tails that may extend to well beyond our photon frequency computational domain, especially for nearly edge-on disks. Finally, our results indicate that increased dissipation in the upper-layers is more important than stresses at the inner disk edge for generating powerful non-thermal emission.

\

Our models also exhibited prominent HFQPOs. Non-zero stresses at the ISCO resulted in sharply increasing effective temperatures as $r/r_g$ decreased, which in turn means most the photons above about $10 \ \rm keV$ came from a narrow range of annuli near the black hole. Compared to our previous work on disks with inner torque, we found QPO profiles with higher quality factors that are more consistent with those obtained from one-zone calculations by \cite{db14}. 

\

While promising, our work likely cannot fully explain the SPL observations where the emission extends to up to MeVs. We found that disks around nearly maximally spinning black holes may have tails that reach hundreds of keVs when seen nearly edge-on, but the same relativistic effects also shift the spectral peak to far higher than $10 \ \rm keV$, which is inconsistent with observations. Ultimately non-thermal electron energy distributions that stretch beyond $100 \ \rm keV$ may be necessary to fully explain the entire sample of high-energy spectral shapes observed in BHBs. 

\

To obtain harder power-law emission that extends to higher energies, we suspect that higher radiative efficiency enhancement (encoded in $\D\ep$) due to stresses at the ISCO would lead to spectral tails that may better match SPL observations over a larger range of dissipation profiles and black hole spins. The resulting annuli atmospheres would be even more effectively thin and have lower surface densities, both of which dramatically increases the difficulty of obtaining converged models. Efforts to overcome these numerical complications are on-going.

\

As mentioned earlier, we adopted modified thin-disk radial solutions \citep{ak00} while radiation pressure dominated flows at near or above the Eddington limit are likely to be geometrically thick. We therefore intend expand this work to slim disk models \citep{ab88}. This would entail first computing radial profiles \citep{sa09a, sa11} that incorporate torques at the inner edge and then using the results as input for vertical structure calculations. \cite{sa09b} explored the effects of different dissipation prescriptions on spectra via ray tracing but did not self-consistently include radiative transfer or consider cases with significant heating near the photospheres. On the other hand, \cite{s11} and \cite{sdm13} used TLUSTY-based spectra as implemented in BHSPEC \citep{dav06} for fitting data but again did not consider the effects of different dissipation profiles.

\

Finally, we will expand this work to incorporate a wider and more fine grained array of dissipation power-law indices $\zeta$. For any annulus vertical structure where the density decreases upwards, dissipation prescriptions of the form given by Equation (\ref{dis2}) would result in peaks away from the mid-plane. We can then, at least qualitatively, model the height dependence of the range of horizontally and time averaged dissipation profiles seen in simulations by adjusting $\zeta$, which physically controls the fraction of accretion power that eventually heats the upper layers. Therefore it is possible to use spectral fitting to constrain the interior distribution of energy dissipation inside observed accretion flows provided that we have a sufficiently high-resolution grid of annuli models that span a large parameter range in $T_{\rm eff}$, $\Sigma_0$, $\O^2$, $\zeta$ and $\D\ep$. Together with SPL slope and QPO measurements, such a `library' of annuli structure and spectra may also enable the extraction $\D\ep$ and hence place approximate quantitative limits of the possible stresses at the ISCO.

\

The authors acknowledge fruitful discussions with O. Blaes, C. Kishimoto, N. Storch and J. Dexter. We also thank the referee for detailed and insightful comments that greatly improved the manuscript. LM and TD were partially supported by a Summer Undergraduate Research Experience grant from the Office of Undergraduate Research at University of San Diego.

\end{document}